\documentclass[reprint,authaffil,TurnOffLineNumbers]{arXiv_mine} 

\usepackage{enumitem}
\usepackage{soul}
\usepackage{todonotes}
\usepackage{multirow}

\begin{document}
\title[Similarity between piano notes: Simulation data]{Perceptual similarity between piano notes:\\ Simulations with a template-based perception model}

%

\author{Alejandro Osses Vecchi}
\affiliation{Human-Technology Interaction group, Department of Industrial Engineering \& Innovation Sciences, Eindhoven University of Technology, 5600MB Eindhoven, the Netherlands,}

\affiliation{Hearing Technology @WAVES, Dept. of Information Technology, Ghent University, Ghent, Belgium\\ \href{mailto:alejandro.osses@ugent.be}{alejandro.osses@ugent.be}, \href{mailto:a.kohlrausch@tue.nl}{a.kohlrausch@tue.nl}}

\author{Armin Kohlrausch}
\affiliation{Human-Technology Interaction group, Department of Industrial Engineering \& Innovation Sciences, Eindhoven University of Technology, 5600MB Eindhoven, the Netherlands,}

\begin{abstract}
In this paper the auditory model developed by \citeauthor{Dau1997b} [J. Acoust. Soc. Am. \textbf{102}, 2892--2905 (1997)] was used to simulate the perceptual similarity between complex sounds. For this purpose, a central processor stage was developed and attached as a back-end module to the auditory model. As complex sounds, a set of recordings of one note played on seven different pianos was used, whose similarity has been recently measured by \citeauthor{Osses2019a} [J. Acoust. Soc. Am. \textbf{146}, 1024--1035 (2019)] using a 3-AFC discrimination task in noise. The auditory model has several processing stages that are to a greater or lesser extent inspired by physiological aspects of the human normal-hearing system. A set of configurable parameters in each stage affects directly the sound (internal) representations that are further processed in the developed central processor. Therefore, a comprehensive review of the model parameters is given, indicating the configuration we chose. This includes an in-depth description of the auditory adaptation stage, the adaptation loops. Simulations of the similarity task were compared with (1) existing experimental data, where they had a moderate to high correlation, and with (2) simulations using an alternative but similar background noise to that of the experiments, which were used to obtain further information about how the participants' responses were weighted as a function of frequency.
\end{abstract}

\maketitle

\section{Introduction}


The sense of hearing provides humans with the possibility to explore and interact with their surrounding sound environment. Examples of this interaction are the ability to localize a sound object, obtain information about its identity, and communicate with others. The ability to access such information by using the hearing system is assumed to be possible due to the existence of internal processes of perceptual organization \cite{McAdams1993}. The information used by these internal processes is sometimes referred to as ``internal representations.'' This term indicates that the hearing system delivers information about the sound object to the brain. The hearing system consists of a mechanical and a neural part. After the mechanical or peripheral auditory processing --that comprises the outer, middle, and inner ear-- the sounds are represented as firing patterns in the auditory nerve. The neural part comprises the connectivity and involved functional mechanisms that transmit the information, i.e., firing patterns of the auditory nerve, through the central nervous system to the brain \cite[e.g.,][]{Kohlrausch2013}.

Although there is consensus that firing patterns at the level of the auditory nerve are encoded according to a frequency-to-place conversion that occurs in the inner ear \cite[e.g.,][]{Robles2001}, there is no similar consensus with respect to higher-level neural processing stages. This has been translated into computational models that consist of stages of peripheral and central processing, with the first following a structure generally based on a tonotopic analysis using a cochlear filter bank and the latter employing diverging schemes to further process simulated firing patterns or, in other words, to obtain and use internal representations. 

\begin{table*}
\caption{Selected list of central processors that are used as back-end stages for published computational models of the auditory periphery. The column ``\# Repr.'' indicates the number of representations required by the ``criterion'' of the central processor.} \vspace{-12pt}
\centering
\begin{ruledtabular}
\begin{tabular}{lcl} 
Central processor type & \# Repr. & Peripheral stage based on \\ \hline
A. Template-based optimal detector \cite{Dau1997b} & 3 & \citet{Dau1997b} \\  
B. Autocorrelator-based pitch analyzer \cite{Meddis1997}  & 1 & \citet{Meddis1991} \\
C. Discriminability analyzer \cite{Fritz2007}             & 2 & \citet{Glasberg2002} \\
D. Envelope analyzer \cite{Joergensen2011}                & 1\footnotemark[1] & \citet{Ewert2000} \\ 
E. Room Acoustic Analyzer \cite{Dorp2013a,Osses2017a,Osses2020a} & 1 & \citet{Breebaart2001a} \\ 
F. Envelope analyzer \cite{Bianchi2019}                   & 1 & \citet{Zilany2009,Zilany2014} \\ 
G. RMS difference detector \cite{Osses2019c,Verhulst2018b} & 3 & \citet{Verhulst2018a} \\ 
H. Template-based discriminability detector \cite{Maxwell2020} & 2 & \citet{Zilany2009,Zilany2014} \\
\end{tabular}
\end{ruledtabular} \vspace{-12pt}
\footnotetext[1]{Processor~D processes ``individual'' speech samples in noise (i.e., one test interval), but the processor also needs to have access to the internal representation of the noise alone in order to generate its output metric.}
\label{tab:central-proc}
\end{table*}

A central processing stage can therefore be seen as a back-end module for the peripheral processing. A selected list of central processors used in published models is presented in Table \ref{tab:central-proc}. A central processor accounts for: (1) high-level neural processing of the hearing system (to a greater or to a lesser extent), and (2) coupling of the internal representation to a certain ``criterion'' (decision stage) that provides concrete information about the processed sound object. In general this latter aspect is assessed by either comparing two or more internal representations (e.g., processors A, C, G, and H in Table~\ref{tab:central-proc}) or by converting the internal representation into a metric believed to reflect some perceptual aspect of the processed sound object (e.g., processors B, D, E, and F in Table~\ref{tab:central-proc}). In this study, a computational model that follows the former rationale is used. We use an updated version of the perception model (PEMO) described by \citet{Dau1997b} with a central processor that compares different internal representations by using the concept of optimal detector. Therefore, our work is concerned with one possible way of comparing internal representations of different sounds. Particularly, the comparison of internal representations is implemented as a three-alternative forced-choice (3-AFC) performance task and it is applied to the evaluation of perceptual similarity between piano note recordings \cite{Osses2019a}. The choice of evaluating piano sounds was motivated by: (1) the complex spectro-temporal properties present in piano sounds, (2) the fact that piano sounds have been thoroughly studied in physical acoustics and we recently quantified differences psychoacoustically \cite{Chaigne2019,Osses2019a}, and (3) the fact that the PEMO model has been primarily applied to study artificial sounds \cite{Dau1996a,Dau1996b,Jepsen2008} and speech \cite{Joergensen2011,Relano-Iborra2019} and less often to other types of sounds, including musical instrument sounds \cite{Huber2006}. Although \citeauthor{Huber2006} applied this auditory model to more diverse sets of sounds, their central processor was adapted to provide a quality metric and, therefore, the goal in their study was to assess judgments of sound quality rather than simulating performance. In this context, the work presented in this study can be seen as an extension to the use of the unified framework offered by the PEMO~model.  

We start by providing an overview of each stage within the PEMO model (Sec.~\ref{sec:model-description}A--B). Subsequently an internal representation obtained from the model is described (Sec.~\ref{sec:intrepr}), introducing the information-based approach we adopted to analyze the contribution from different frequency regions in the internal representations of our stimuli. In Sec.~\ref{sec:methods}, the dataset of piano sounds and background noises we used are presented together with the adopted simulation protocol. The obtained results are presented in Sec.~\ref{sec:results} and discussed in Sec.~\ref{sec:discussion}. This includes simulations using the same noises as used in the reference listening experiments, and using an alternative noise that has slightly different spectral properties. These latter simulations were used to provide insights into the weighting of frequency information used by the listeners during the reference experiment. We conclude this paper by providing perspectives on the use of the PEMO model for applications different from those for which the model was developed.

\section{Model description} 
\label{sec:model-description}

\begin{figure*}
	\centering
	\includegraphics[width=\textwidth]{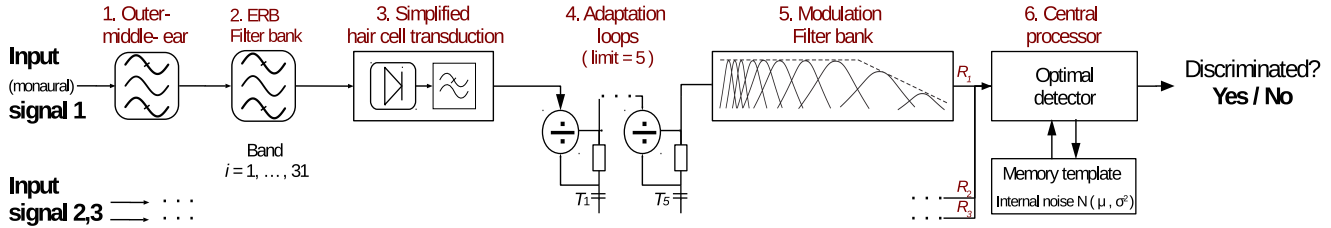} 
	\caption{Block diagram of the perception model (PEMO) model. Each of its stages is explained in the text.}
	\label{fig:auditory-model}
\end{figure*}

The block diagram of the perception model \cite[PEMO][]{Dau1997b} is shown in Fig.~\ref{fig:auditory-model}. Each of the model stages is described in this section.\footnote{A similar but much shorter description of the stages of the PEMO model, including the adopted central processor, can be found in our previous simulation paper \cite{Osses2018a}.} These descriptions were compiled from previous published implementations of the PEMO model \cite{Dau1996a, Dau1997b, Verhey1999, Derleth2000} or variants of it \cite{Breebaart2001a, Breebaart2001b, Breebaart2001c, Jepsen2008}. The current model configuration has been included in the AMT toolbox \cite{Soendergaard2013}, v0.10 (Appendix~\ref{app:PEMO-in-AMT}).

\subsection{Processing stages}
\label{sec:model-stages}

\subsubsection{Outer- and middle-ear filtering}

This stage accounts for the effects of outer and middle ear on the incoming signal and it is implemented as two cascaded 512-tap finite impulse response (FIR) filters. The outer-ear filter introduces a transfer function from headphones to the tympanic membrane, emphasizing frequencies around 2750~Hz and attenuating frequencies above 6000~Hz \cite[][]{Pralong1996}. The middle-ear filter introduces a transfer function from the tympanic membrane to the stapes, approximating the (peak-to-peak) velocity of the stapes in response to pure tones. The filter acts as a band-pass filter (BPF) with a maximum at 800 Hz (unit gain, $0$ dB) and slopes of about 6~dB/octave below and above that frequency \cite{Lopez-Poveda2001,Goode1994}. The resulting frequency response shows two local peaks at 2750 Hz ($-3$~dB gain) and 5000 Hz ($-13$ dB). This stage was included in the auditory model of \citet{Jepsen2008} but not in previous versions of the PEMO model. 

\subsubsection{Gammatone filter bank}

This set of filters corresponds to a level-independent approximation of a critical-band filter bank. The Gammatone filter bank consists of 31 bands having center frequencies between 87~Hz (3 ERB$_N$) and 7819~Hz (33~ERB$_N$), spaced at 1 ERB. The model uses band-limited signals obtained from the real part of the complex valued all-pole implementation described by \citet{Hohmann2002}. All further processing stages of the model work independently on each auditory filter output.

\subsubsection{Hair-cell transduction}

This stage accounts for a simplified inner hair cell processing. It simulates the transformation from mechanical oscillations in the basilar membrane into receptor potentials in the inner hair cells. The signals are first half-wave rectified and then low-pass filtered using five first-order infinite impulse response (IIR) filters with a cut-off frequency (f\textsubscript{cut-off}) of 2000~Hz. The combined effect of the cascaded low-pass filters (LPFs) is equivalent to applying a fifth-order IIR filter with a f\textsubscript{cut-off} of 770~Hz. Therefore, frequency components below 770~Hz are almost unaffected, preserving the phase information (maximum attenuation of 3~dB at 770~Hz). Frequency components between 770 and 2000~Hz are gradually attenuated (attenuations between 3 and 15~dB, respectively), meaning that the phase information is gradually lost. For frequency components above 2000~Hz almost all phase information is removed (attenuation greater than 15~dB, slope of $-30$~dB/octave). This way of removing phase information is consistent with the decrease of phase locking observed in the auditory nerve \cite{Breebaart2001a}.

\subsubsection{Adaptation}
\label{sec:adaptation}

This stage simulates the non-linear adaptive properties of the hearing system at the level of the auditory nerve, where changes in the transformation characteristic (system gain) are introduced according to the input signal level \cite[e.g.,][]{Kohlrausch1992}. If the change of the input level is rapid compared to a time constant $\tau$, the system gain increases transforming the levels linearly. For slower variations, the system gain gradually decreases and the input levels are compressed. The adaptation loops structure is used for this purpose \cite{Pueschel1988, Muenkner1993, Dau1996a}, which consists of 5~feedback loops, each of them having a different time constant ($\tau$=5, 50, 129, 253, 500~ms). A detailed description of this stage is given in Appendix~\ref{app:adaptation} \cite[see also][his App.~C]{Osses2018}. In short, each loop acts as a low-pass filter that gets charged or discharged (applying more or less compression, respectively) depending on the instantaneous characteristics of the incoming signals and the corresponding $\tau$. In line with previous model implementations, we introduced an overshoot limitation, meaning that the (output) amplitudes for rapid input changes (relative to $\tau$) are limited. However, unlike previous studies, the limiter factor was set to a lower value (lim=$5$ instead of 10) which lead to a more severe compression of overshoot responses. Due to the relevance of the note onset in piano sounds, the choice of this new limiter factor strongly influenced the simulations that are shown later in this paper. For this reason, the effect of using the new limiter factor is described in Sec.~\ref{sec:intrepr}.

\subsubsection{Modulation filter bank}

The modulation filter bank processes the incoming signal in terms of changes in its envelope. In this stage the same implementation as suggested by \citet{Jepsen2008} is used. First, a reduction in the sensitivity to modulation frequencies above 150~Hz is introduced \cite{Kohlrausch2000} by applying a LPF (f$\textsubscript{cut-off}=150$~Hz, roll-off$=6$~dB/octave). The filter bank comprises a maximum of 12 filters (Table \ref{tab:mfb}) that have two different envelope frequency domains: (1) Bands 1 to 3 (mf$_c\leq$10~Hz, constant bandwidth of 5.4~Hz, Table~\ref{tab:mfb}) where the real-valued part of the filtered (band-limited) signals is used. This processing preserves the modulation phase information; (2)~Bands 4 to 12 (mf$_c>$10~Hz, constant quality factor Q=2\footnote{Other studies that use modulation filters have used alternatively wider filters, with Q factors of 1 \cite{Ewert2000, Nelson2004, Joergensen2011, Wallaert2017}. We did not investigate the effect of filter tuning.}) where the norm of the complex signals is used, which represents an approximation to the Hilbert envelope \cite{Hohmann2002}. Although this process reduces considerably the modulation phase information, the modulation energy within the respective band is maintained. An attenuation factor of $\sqrt{2}$ is applied to the resulting signals \cite{Jepsen2008}.

The modulation filters for each audio frequency band are limited to filters having an mf$_c$ below a quarter of the audio center frequency $f_c$. This is motivated by results presented by \citet{Langner1988}, where evidence is provided that neural activity in the ascending auditory path (auditory brainstem) has best modulation frequencies limited to that frequency range (i.e., mf$_c<f_c/4$). The modulation filter bank output represents an internal representation $R_x$ that approximates the time-frequency modulation sensitivity at the level of the inferior colliculus \cite{Dau1997b}. The amplitudes of the representation $R_x$ are scaled in arbitrary or model units (MU) \cite[][]{Kohlrausch1992,Dau1996a}.

\begin{table}
\centering
\caption{Impulse-response-derived parameters of the modulation filter bank,\textsuperscript{a} where mf$_c$ is the center frequency of each filter, $f\textsubscript{inf}$ and $f\textsubscript{sup}$ are the corresponding low and high cut-off frequencies ($-3$ dB points), BW represents the bandwidth ($f\textsubscript{sup}-f\textsubscript{inf}$), Q the quality factor (mf$_c$/BW), and $f_c$ indicates the center frequency of the audio bands such that mf$_c<f_c/4$.} \vspace{-12pt}
\begin{ruledtabular}
\begin{tabular}{crcrccll}
Band & \multicolumn{2}{c}{Frequency [Hz]} & BW  &  & audio-band f$_c$ \\
Nr. & mf$_c$ & $f$\textsubscript{inf} - $f$\textsubscript{sup} & [Hz] & Q & [ERB$_N$] \\
\hline
1  &  2.7 &   0.0 -   2.7 &    2.7 & -\footnotemark[2]   & 3-33 \\
2  &  5.0 &   2.7 -   8.1 &    5.4 & 0.9 & 3-33 \\
3  &  10.0 &   7.4 -  12.8 &   5.4 & 1.9 & 3-33 \\
4  &  16.7 &  12.8 -  20.9 &   8.1 & 2.1 & 3-33 \\
5  &  27.8 &  20.9 -  35.0 &  14.1 & 2.0 & 4-33 \\
6  &  46.3 &  35.0 -  58.5 &  23.6 & 2.0 & 4-33 \\
7  &  77.2 &  57.9 -  96.9 &  39.0 & 2.0& 8-33 \\
8  & 128.6 &  96.9 - 160.8 &  63.9 & 2.0& 11-33 \\
9  & 214.3 & 160.8 - 268.5 & 107.7 & 2.0& 15-33 \\
10 & 357.2 & 268.5 - 446.8 & 178.3 & 2.0& 19-33 \\
11 & 595.4 & 446.8 - 744.2 & 297.4 & 2.0& 23-33 \\
12 & 992.3 & 744.2 -1240.9 & 496.6 & 2.0& 27-33
\end{tabular}
\end{ruledtabular}\vspace{-12pt}
\footnotetext[1]{The 150-Hz LPF was omitted for this analysis.}
\footnotetext[2]{Band 1 is a LPF, for which no Q factor is indicated.}
\label{tab:mfb}
\end{table}

\subsubsection{Central processor}
\label{sec:central-proc}

In this stage, the internal representation $R_x$ of each interval ($x$=1, 2, 3) is compared with a reference representation or ``sound image'' that is stored in the memory (top-down component) of the model --the template $T_p$-- to mimic the decision that human listeners perform in a 3-AFC task. This central processor is inspired by the concept of an optimal detector \cite[][Ch. 6 and 7]{Green1966}. The model is hence used as an artificial listener,\footnote{In the literature (and in this paper), the terms ``artificial listener'' and ``artificial observer'' are used interchangeably.} where the template corresponds to an expected sound representation that gives a clear indication to the artificial listener about ``what to listen for'' \cite{Dau1996a}. For this reason the template should be derived in a condition that is easily detectable, i.e., in a supra-threshold condition. For detection and discrimination tasks conducted in noise this is at a high signal-to-noise ratio (SNR). In our simulations we adopted an SNR\textsubscript{supra} equal to 21~dB, which represents an SNR that is 5~dB above the initial SNR used to collect the reference (experimental) data.

\paragraph{Use of a template}\mbox{}

In order to obtain a decision outcome using the internal representations $R_x$ and the stored template $T_p$, the representation $R_x$ ($x$=1, 2, 3) that has the highest similarity with the template $T_p$ is indicated by the artificial listener as containing the target interval. The cross-correlation value (CCV) can be used for this purpose:
\begin{equation}
	\mbox{CCV}_x=\frac{1}{f_s} \sum_{n=1}^N \Delta R_x[n]\cdot T_p[n] 
	\label{eq:CCV}
\end{equation}

\noindent where $\Delta R_x$ and $T_p$ are digital signals with sampling rate $f_s$ and $N$ samples. The difference representation $\Delta R_x$ is obtained from the piano-plus-noise representation $R_x$ and the corresponding noise alone representation $R_{N,x}$, similar to how it has been done in previous tone-in-noise simulations \cite{Dau1996a}. However, as explained next, our similarity task further required to enable the central processor to use two templates.

\paragraph{Template in a similarity task}\mbox{}
\label{sec:template}

\begin{figure}
	\centering
\includegraphics[width=0.47\textwidth]{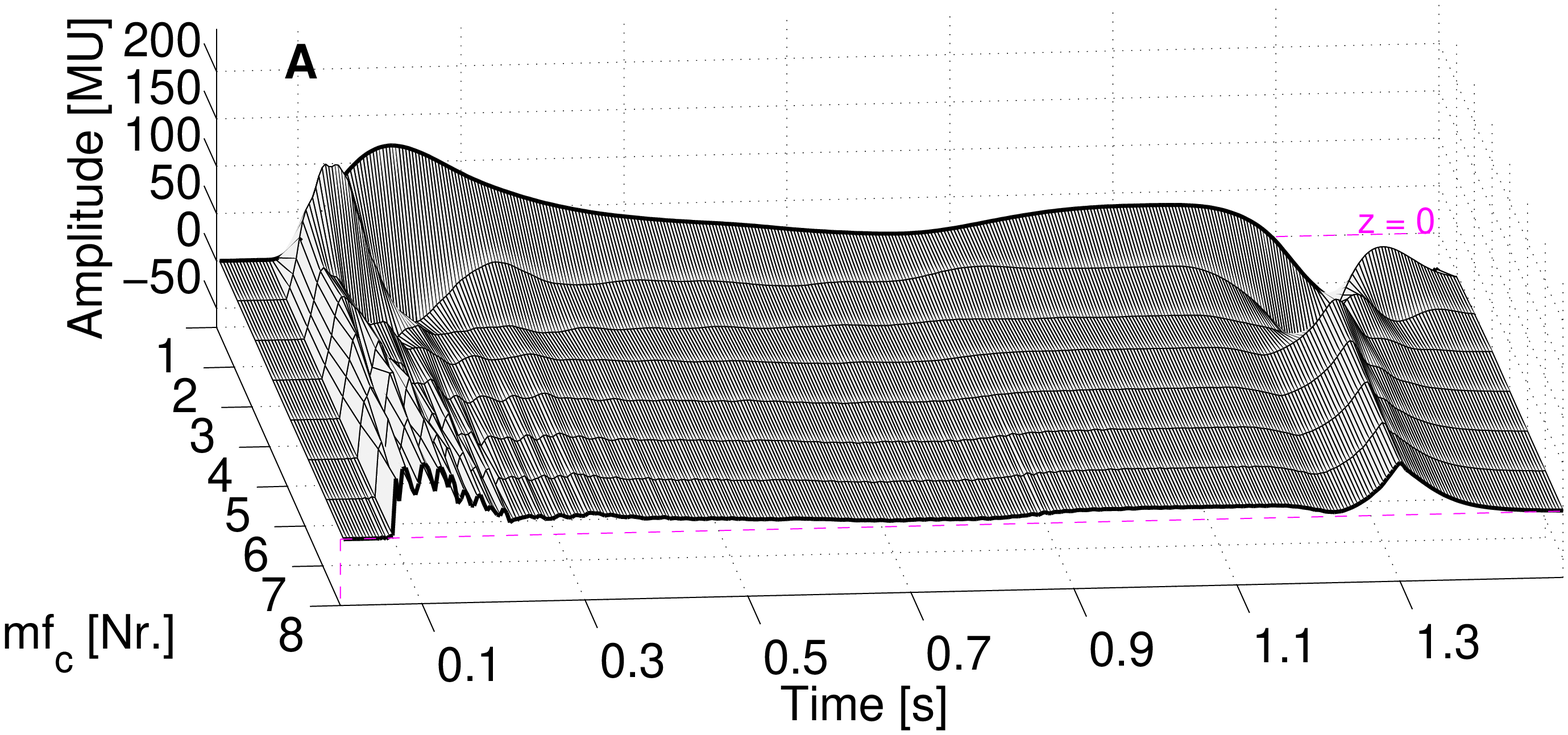}
\includegraphics[width=0.47\textwidth]{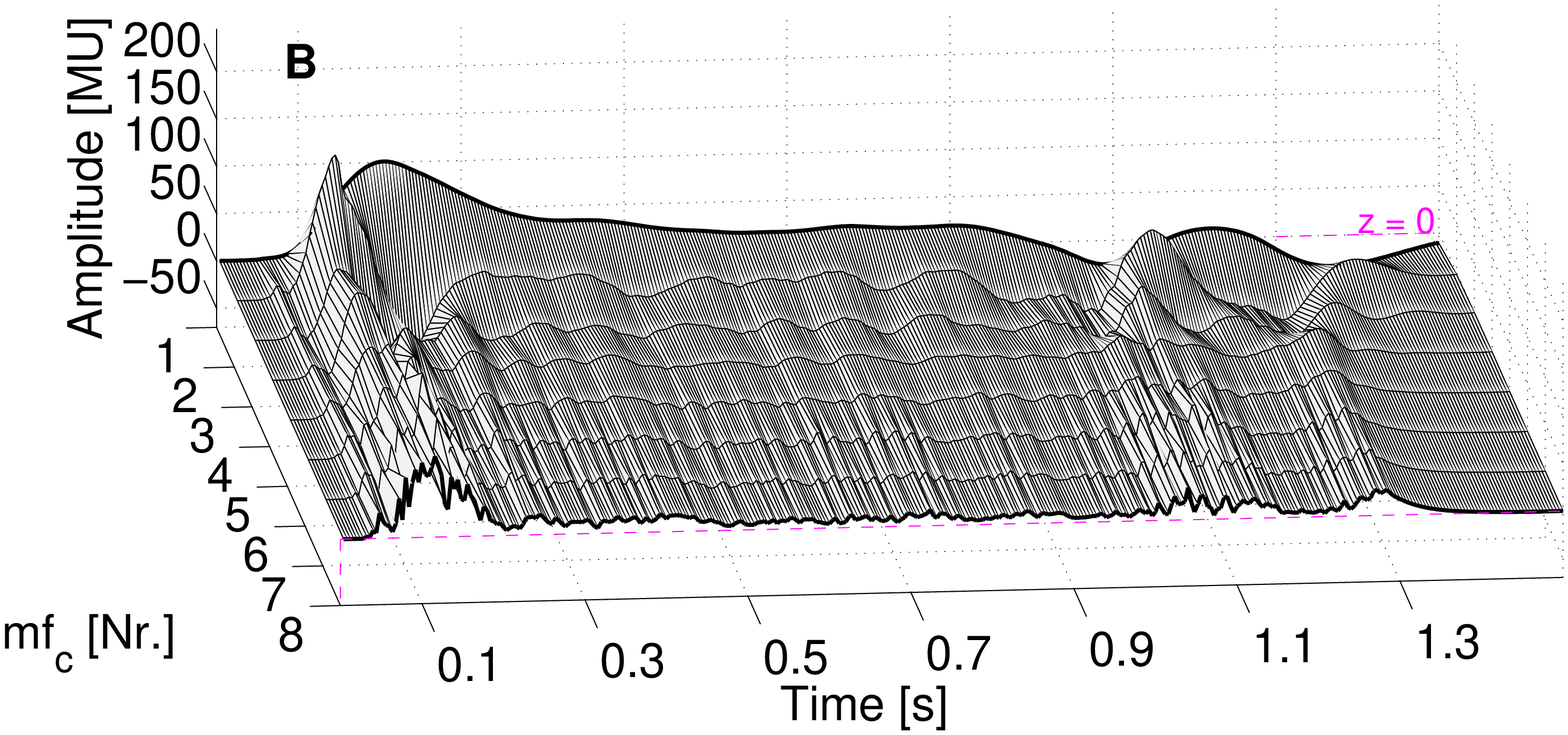}
	\caption{Internal representation for the recordings of pianos P1 (panel \textbf{A}) and P3 (panel \textbf{B}). These internal representations correspond to the outputs of the peripheral stage of the PEMO model. For clarity, the analysis of only one audio frequency band ($f_c=520$~Hz) is shown. This band has 8 modulation filters with frequencies mf$_c$ between 1.4 and 128.6~Hz.}
	\label{fig:intrep-pianos}
\end{figure}

The template we adopted to simulate the 3-AFC similarity task described by \citet{Osses2019a} is determined by: (a) the target and ``reference'' sounds, and; (b) two or more realizations of a noise that can efficiently mask the properties of both piano sounds, given that the 3-AFC task is implemented as a piano-in-noise discrimination. To account for the latter aspect, noises that are generated using a modified ICRA algorithm\footnote{ICRA stands for International Collegium of Rehabilitative Audiology. We decided to keep this denomination in the current study, given that our background noises had the same objective than the original ICRA noises \cite{Dreschler2001} of creating noises with the same spectro-temporal properties as the input sounds.} \cite[version~A,][]{Osses2019a} are used in every piano presentation. For the first aspect, the template $T_p$ is derived from the internal representations of both, the target piano $R_t$ and reference piano $R_r$, because their discrimination threshold depends on how different they are from each other. In the course of this study different ways of deriving $T_p$ using $R_t$ and $R_r$ were evaluated. Only the adopted approach is described in this section. The interested reader is referred to \cite[][his App.~E]{Osses2018} where two other (discarded) template approaches are described. 


In the adopted approach, two templates are derived, $T_{p,t}$ and $T_{p,r}$, for the target and reference piano sounds, respectively. For each template, an average representation of the piano sounds embedded in four different ICRA noise realizations at a highly discriminable condition (SNR\textsubscript{supra}$=21$~dB) is obtained.\footnote{The number of ICRA noise realizations (four) used to derive each average piano-plus-noise representation was an arbitrary choice.} The templates are normalized to unit energy \cite{Dau1996a} to satisfy:
\begin{equation}
	E_t=\frac{1}{f_s} \sum_{n=1}^N T^2_{p,t}[n]=1 \mbox{,\hspace{15pt}} E_r=\frac{1}{f_s} \sum_{n=1}^N T^2_{p,r}[n]=1
	\label{eq:E-here}
\end{equation}

\noindent where $N$ corresponds to the number of samples used by the artificial listener to make the decision,\footnote{In analogy to the theory of optimal detectors presented by \citet{Green1966}, we treat the templates $T_{p,t}$ and $T_{p,r}$ as ``expected signals'' along one (temporal) dimension. In fact, there are two other dimensions: audio and modulation frequency. Considering all template dimensions and following the nomenclature of Eq.~\ref{eq:Itot}, Eq.~\ref{eq:E-here} would turn into $E_t=\frac{1}{f_s} \sum_{m=1}^M \sum_{k=1}^K \sum_{n=1}^N T^2_{p,t\mbox{ }mk}[n]=1$.} during what we refer to as the observation (listening) period $t\textsubscript{obs}$.

\paragraph{Use of two templates}\mbox{}

During the simulations, the templates $T_{p,t}$ and $T_{r,t}$ are compared with the $x$-intervals in each 3-AFC trial. Equation~\ref{eq:CCV} is used for this purpose, which uses a difference representation $\Delta R_x$ obtained from $R_x$ and the corresponding noise representation ($R_{N,x}$) at the SNR of the trial.\footnote{The use of difference representations $\Delta R_x$ is relevant to obtain CCV$_{x,t}$ and CCV$_{x,r}$ values that are in a comparable range. This ``CCV baseline'' is needed because the unit energy normalization of the templates is done independently and is, therefore, an inherent problem of the use of two templates. Subtracting the noise alone representations ($R_{N,x}$) in the CCV calculation implies that the resulting CCV$_{x,t}$ and CCV$_{x,r}$ values indicate the contribution of information of piano $x$ relative to that of noise $x$.} Six CCV values are obtained: 
\begin{eqnarray}
	CCV_{x,t}=\frac{1}{f_s} \sum_{n=1}^N \Delta R_x[n]\cdot T_{p,t}[n], & \mbox{\hspace{12pt}with }x=1,2,3 \nonumber \\ 
	CCV_{x,r}=\frac{1}{f_s} \sum_{n=1}^N \Delta R_x[n]\cdot T_{p,r}[n] &  
	\label{eq:CCV-with-temp}
\end{eqnarray}

\noindent Based on these values, the artificial listener chooses the interval that is more likely to contain the target sound. If we assume that the target interval is presented in the first interval ($x=1$), then for a correct discrimination:

\begin{eqnarray}
\max{\left\{ \widehat{CCV}_{x,t}\right\} }&=&\widehat{CCV}_{1,t}, \mbox{\hspace{12pt}and } \nonumber\\ 
\min{\left\{ \widehat{CCV}_{x,r}\right\} }&=&\widehat{CCV}_{1,r}
\label{eq:CCV-criterion} 
\end{eqnarray}

\noindent In other words, the target template $T_{p,t}$ is expected to elicit the maximum CCV value when being correlated with the target interval. Likewise, the reference template $T_{p,r}$ elicits higher CCV values when being correlated with the reference intervals and therefore the lowest CCV value is attributed to the target interval. The hat symbol indicates that the CCV values differ from the exact definition given in Eq.~\ref{eq:CCV-with-temp}. This is caused by an internal noise, whose values are drawn from a Gaussian distribution $N(\mu,\sigma^2)$ with mean $\mu=0$ and standard deviation $\sigma$=$10.1$~MU. In our implementation of the internal noise, one independent number is added for each CCV$_x$ value:
\begin{eqnarray}
	\widehat{CCV}_{x,t} &=& CCV_{x,t}+N_x(\mu,\sigma^2) \nonumber \\ 
	\widehat{CCV}_{x,r} &=& CCV_{x,r}+N_x(\mu,\sigma^2) 
	\label{eq:CCV-with-temp-hat}
\end{eqnarray}

\noindent Since $\mu=0$, the standard deviation $\sigma$ corresponds to the actual source of internal variability in the decision process. The use of this Gaussian noise leads to a reduction in the process performance when either the CCV$_{x,t}$ values get close to each other or when the CCV$_{x,r}$ values do. The standard deviation $\sigma=10.1$~MU was obtained by running an increment-discrimination task with each piano sound and tracking the amount of noise needed to produce an average performance of 70.7\% for a difference in level of $\Delta L=1$~dB \cite[][his App.~D]{Osses2018}.

\paragraph{Alignment between piano representations}\mbox{}

\begin{figure}
	\centering
\includegraphics[width=0.47\textwidth]{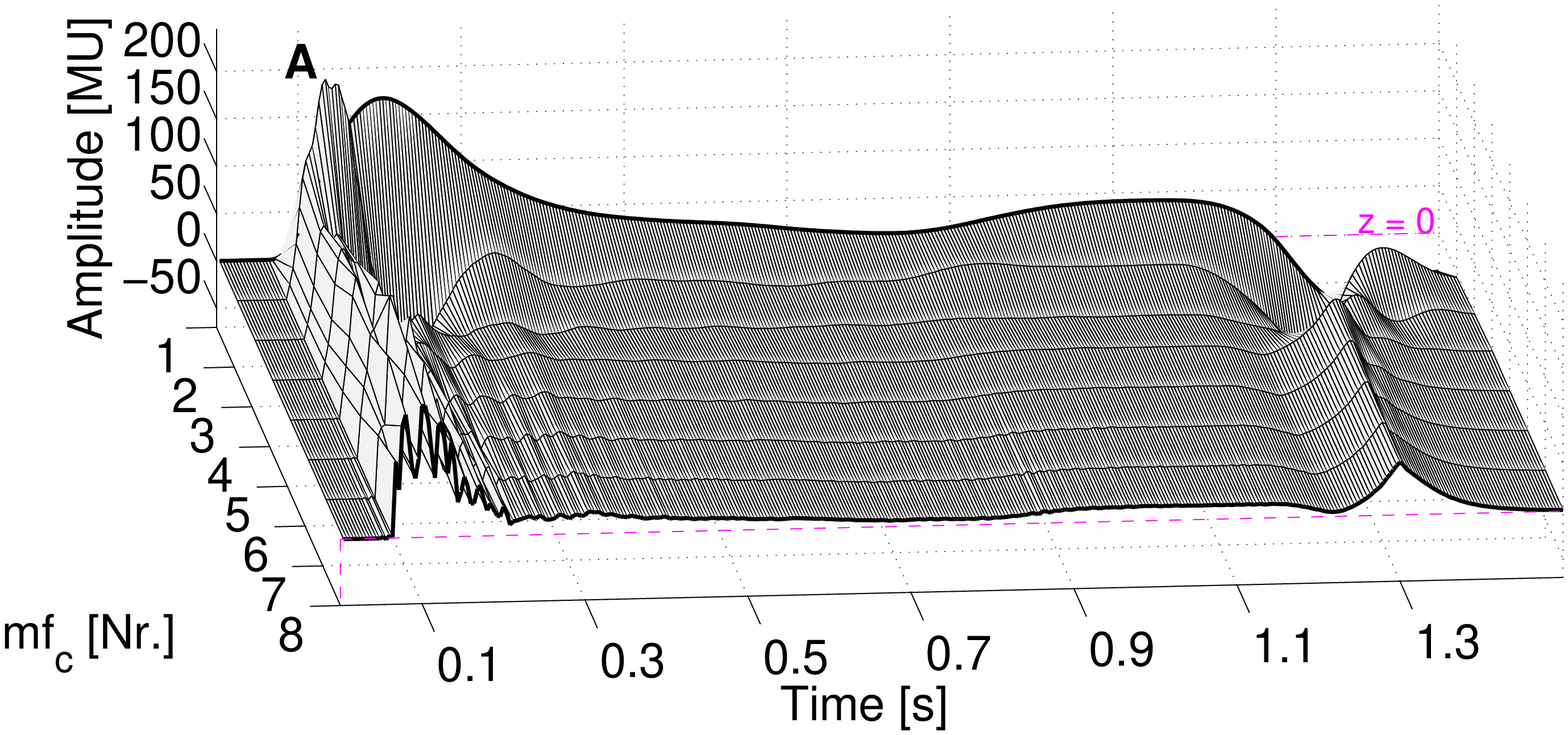}
\includegraphics[width=0.47\textwidth]{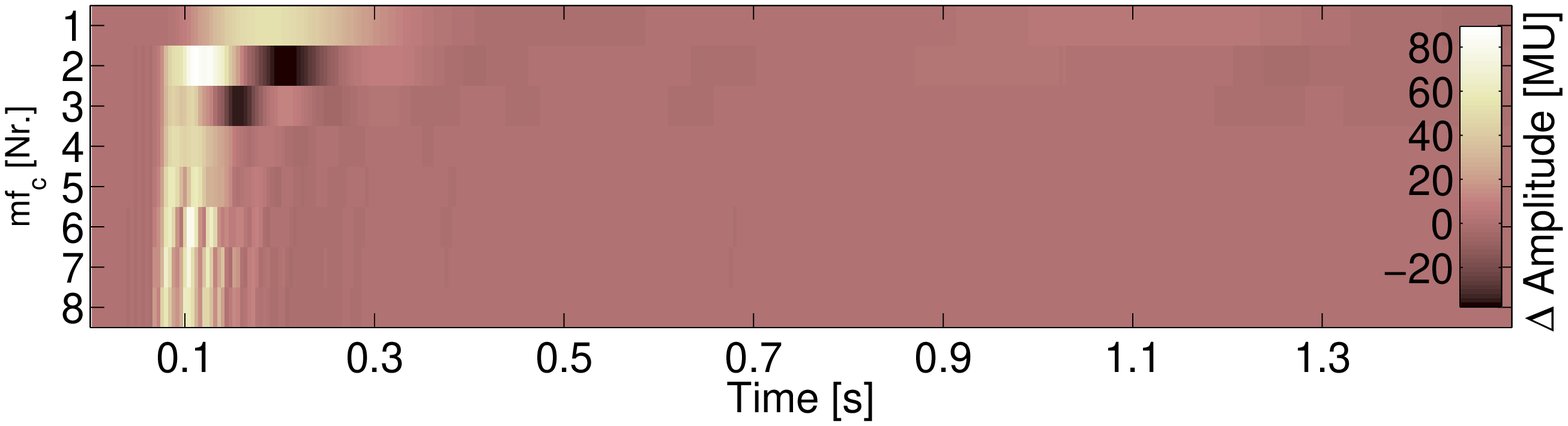} 
	\caption{(\textbf{A}) Internal representation of piano P1 with a factor lim=$10$ in the adaptation loops stage for the audio band centered at $f_c=520$~Hz. This representation has overall larger amplitudes than the representation shown in Fig.~\ref{fig:intrep-pianos}A (with lim=$5$).  (\textbf{B}) Difference between the P1 representations with limiter factors of 10 (panel A) and 5 (Fig.~\ref{fig:intrep-pianos}A). The difference is more prominent around the sound onset with higher amplitudes of up to 80.6 MU (modulation band~2) for the representation with lim=$10$ (light areas). As a consequence of the higher overshoot amplitudes for lim=$10$, more compression is applied immediately after the overshoot peaks. This is indicated by the dark regions in band 2 and~3.}
	\label{fig:P1-lim-5-and-10}
\end{figure}

A final aspect that was considered in the decision criterion is that the cross-correlation between templates and interval representations should deliver the highest possible CCV values. As described in Sec.~\ref{sec:piano-stimuli}, our piano stimuli are aligned to have the note onset at a time stamp of 0.1~s. This alignment seemed to be sufficient for listeners to perceive aligned piano-plus-noise intervals during the experimental sessions \cite{Osses2019a}. However, this did not always ensure a maximum CCV value in the simulated decision process, especially when correlating the target piano representation with the reference template $T_{p,r}$ or the reference piano representation with the target template $T_{p,t}$. The workaround we implemented was the assessment of the cross-correlation function between each template and interval for time lags between $-50$~ms and $50$~ms, with 1-ms steps. In this way, the CCV$_x$ values of Eq.~\ref{eq:CCV-with-temp-hat} are obtained from the maximum of the corresponding cross-correlation function. Another interesting approach that may have produced  similar maximization results is the time stretching between representations \cite{Agus2012}.

\begin{figure*}
	\centering
\includegraphics[height=0.18\textheight]{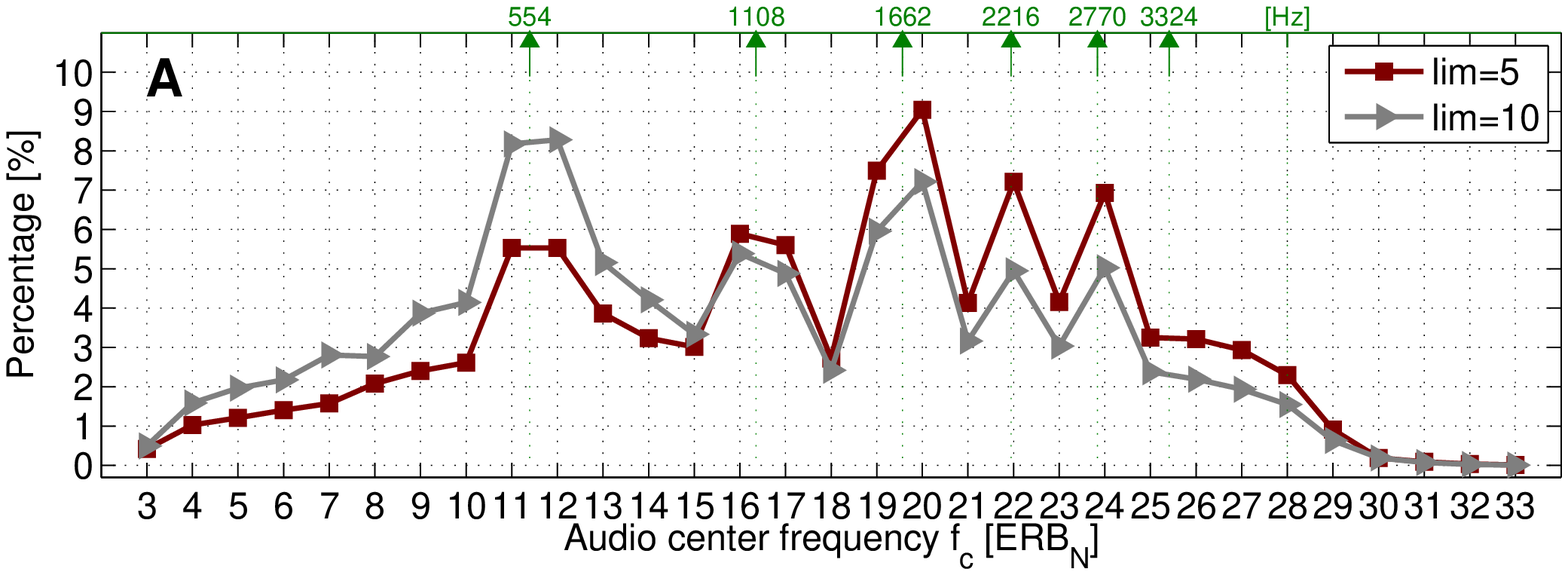}
\includegraphics[height=0.18\textheight,clip=true,trim=.75cm 0 0 0]{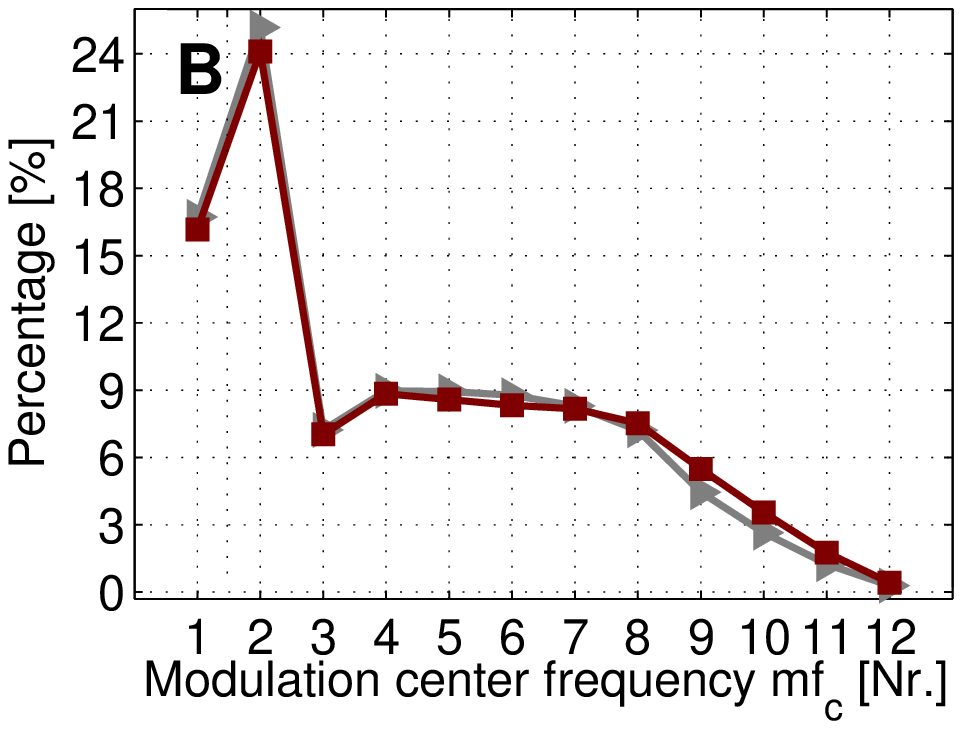}
	\caption{Information in the internal representation of piano P1 for each audio frequency band ($I_m/I\textsubscript{tot}$, panel \textbf{A}), and modulation frequency band ($I_k/I\textsubscript{tot}$, panel \textbf{B}). The maroon square markers indicate the information in the representation with lim=$5$. The gray triangle markers indicate the information in the representation with lim=$10$. The values per band are expressed as percentage with respect to the total information $I\textsubscript{tot}$. The points along the ERB scale that correspond to $f_0$=$554$~Hz and its first harmonics are indicated by the green labels on the top axis.} 
	\label{fig:energy-per-band}
\end{figure*}

\subsection{Sources of internal and external variability}
\label{sec:sources}

The perception of specific sounds is influenced by uncertainties in the stimuli and by internal variability caused, e.g., by imperfections in memory and changes in the concentration level \cite[][]{Yost1989}. In this study we differentiate between sources of variability that are internal or external. Uncertainties in the stimuli are related to an external source of variability, while the effects of human memory and concentration correspond to sources of internal variability. To (partly) account for variations in listening performance due to sources of internal variability, an internal noise is often used in models of auditory processing, which in the PEMO model is simulated as an additive Gaussian noise (Eq.~\ref{eq:CCV-with-temp-hat}). In threshold-detection tasks in background noise, a typical source of external variability is the use of running noises, i.e., the use of different noise relizations that have the same statistical properties within and between trials \cite[e.g.,][]{Klitzing1994}. In the instrument-in-noise test, a running noise condition is approximated by using 12 different ICRA noise realizations for each piano pair \cite{Osses2018a,Osses2019a}. Another source of external variability in the test is the presentation level of each interval, which is randomized (roved) by levels uniformly distributed in the range $\pm 4$~dB.

\subsection{Description of internal representations}
\label{sec:intrepr}

The internal representations of this study were simulated using the PEMO model with an adaptation stage that had a stronger limiter factor (lim=5) than typically used in the literature (lim=10). For this reason this section shows how this parameter choice influenced the piano representations that were later used to simulate the discrimination thresholds between pianos.

\subsubsection{General description of the representations}

The internal representation of pianos P1 and P3 (see Sec.~\ref{sec:piano-stimuli}) after the modulation filter bank (output of Stage~5, Fig.~\ref{fig:auditory-model}) is shown in Fig.~\ref{fig:intrep-pianos} for the frequency band that contains the fundamental frequency $f_0$ of the piano note ($f_c$=$11$ ERB$_N$ or $520$~Hz, $f_0$=554~Hz). The piano sounds start at $t=0.1$~s and their onsets occur shortly thereafter. The onset of the lowest modulation filter (band~1, mf$_c$=$2.7$~Hz, see Table~\ref{tab:mfb}) occurs approximately at $t$=$0.20$~s, for band~2 at $t$=$0.15$~s and for the rest of the bands between $t$=0.10 and $t$=0.11~s. It can also be observed in the figure that after the piano onset, the amplitudes in bands~2 to 8 of P3  present more variations than those of P1, especially between $t$=1 and $1.3$~s.

\subsubsection{Effect of the stronger limiter factor}


The effect of using a stronger limiter factor in the adaptation loops (Stage~4, Fig.~\ref{fig:auditory-model}) is illustrated for one of the piano representations (piano P1). The following description is  qualitatively valid also for the other 6 piano sounds of the dataset (not shown here). 

Two configurations of the adaptation loops are used: Using a limiter factor lim=$5$ (as used in this study, Fig.~\ref{fig:intrep-pianos}\textbf{A}) and using a factor lim=$10$ (as used in the literature, Fig.~\ref{fig:P1-lim-5-and-10}\textbf{A}). The representation with lim=$5$ has amplitudes that range between $-27$ and 142~MU. The amplitudes of the representation with lim=$10$ range between $-62.5$ and 231.5~MU. In both cases the minimum and maximum amplitudes occur in band~2 (centered at mf$_c$=5~Hz). The difference between both representations is shown in Fig.~\ref{fig:P1-lim-5-and-10}\textbf{B}. This difference is more prominent around the sound onset, visible as light areas in Fig.~\ref{fig:P1-lim-5-and-10}\textbf{B}, indicating less compressed amplitudes for the representation with lim$=10$ than those of the representation with lim=$5$. This is followed, however, by a short but strong compression immediately after the overshoot peaks in the representation with lim=$10$, which is indicated by the dark regions in the figure. This  compression is at most $37.9$~MU (band~2) below the P1 representation with lim=$5$. The largest difference between representations is found in band~2, where the representation with lim=$10$ reaches an amplitude 89.5~MU above the maximum of the representation with lim=$5$. 

\subsubsection{Information in the internal representations}

The internal representations obtained with the PEMO model have three dimensions: time ($n$), audio frequency ($m$), and modulation frequency ($k$). Here we suggest an approach to benefit from the information available across dimensions within a representation, that can be used to compare between two (or more) representations. We illustrate this method for the comparison of P1 representations using lim=5 and lim=10.

The contribution of information for each audio and modulation frequency can  be assessed using:
\begin{equation}
I_m=1/f_s \cdot \sum_{k=1}^K \sum_{n=1}^N R_{mk}^2[n] \nonumber
\end{equation}

\begin{equation}
I_k=1/f_s \cdot \sum_{m=1}^M \sum_{n=1}^N R_{mk}^2[n] \label{eq:Ik}
\end{equation}

\noindent This expression is similar to Eq.~\ref{eq:CCV-with-temp}, but the subindexes $m$ and $k$ have been added to indicate that the ``integration of information'' can be done by either deriving the contribution (1) $I_m$ of $M=31$ audio frequency bands across all modulation filter bands, or (2) $I_k$ of $K=12$ modulation frequency bands across all audio frequency bands. The contributions $I_m$ and $I_k$ can be expressed as percentages of the total information $I\textsubscript{tot}$, with $I\textsubscript{tot}$ given~by:
\begin{equation}
	I\textsubscript{tot}=\sum_{m=1}^M I_m = \sum_{k=1}^K I_k = 1/f_s \cdot \sum_{m=1}^M\sum_{k=1}^K\sum_{n=1}^N R_{mk}^2[n]
	\label{eq:Itot}
\end{equation}
\noindent The results of this information-based analysis in the comparison of P1 representations with lim=5 and lim=10 are shown in Fig.~\ref{fig:energy-per-band}\textbf{A}--\textbf{B} for the audio ($I_m/I\textsubscript{tot}$) and modulation bands ($I_k/I\textsubscript{tot}$), respectively. It can be observed that the use of a stronger limitation (lim=5) increases the relative contribution of higher audio frequency bands (Fig.~\ref{fig:energy-per-band}\textbf{A}), while no substantial change in the information weighting is observed across modulation bands (Fig.~\ref{fig:energy-per-band}\textbf{B}).  For the representation with lim=$5$, the audio frequency bands with $f_c$ below $15$~ERB$_N$  ($f_c<924$~Hz) comprise only  30.9\% of the information in contrast to 45.6\% for the representation with lim=$10$ in the same frequency region. In terms of modulation frequency content, which is similar for both representations, bands 1 and 2 (mf$_c\leq5$~Hz) comprise about 40\% of the information and the remaining 60\% is distributed across bands~3 to~12. 

\section{Methods}
\label{sec:methods}

\begin{figure}
\centering
\includegraphics[width=0.4\textwidth]{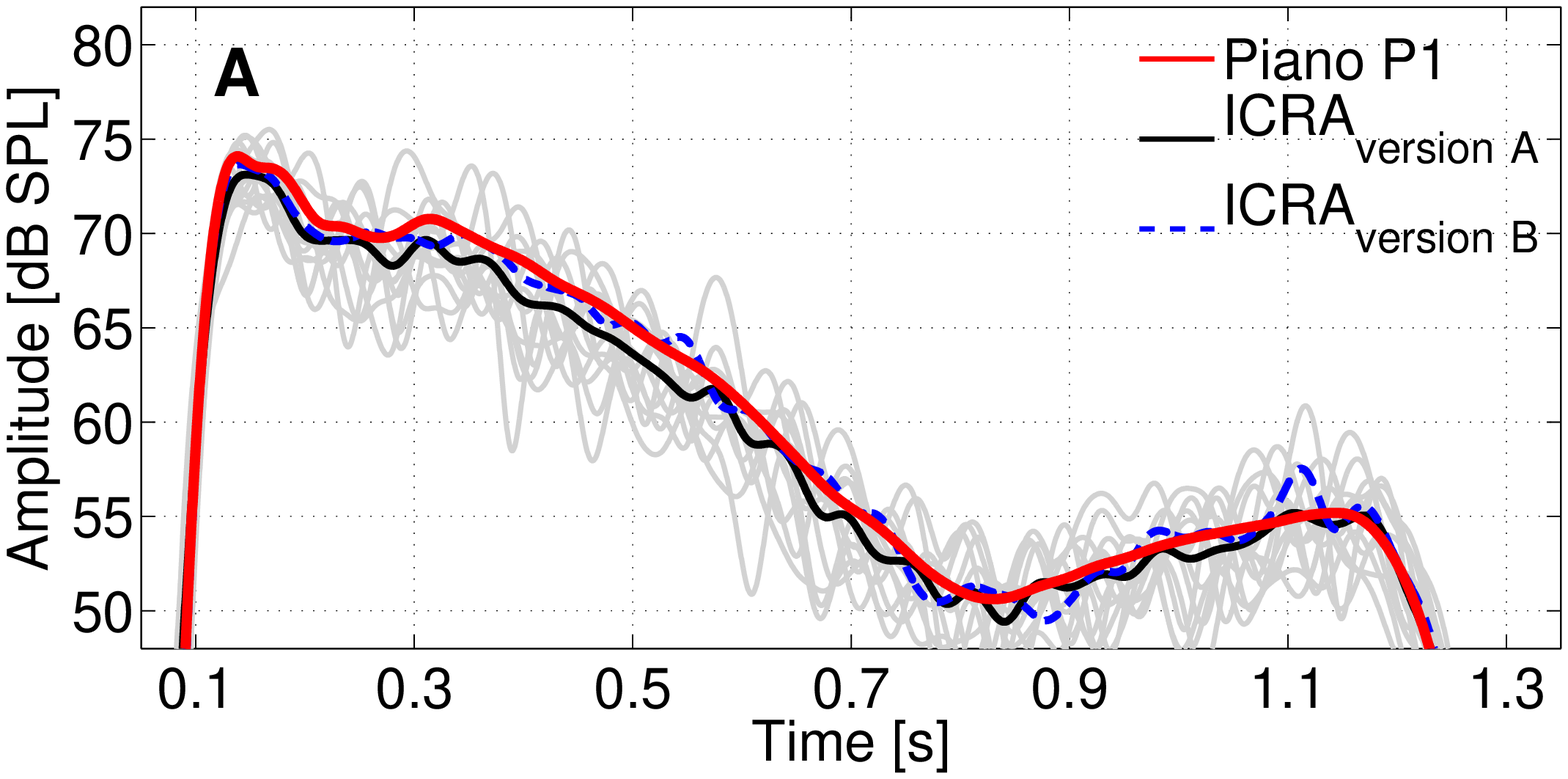}
\includegraphics[width=0.4\textwidth]{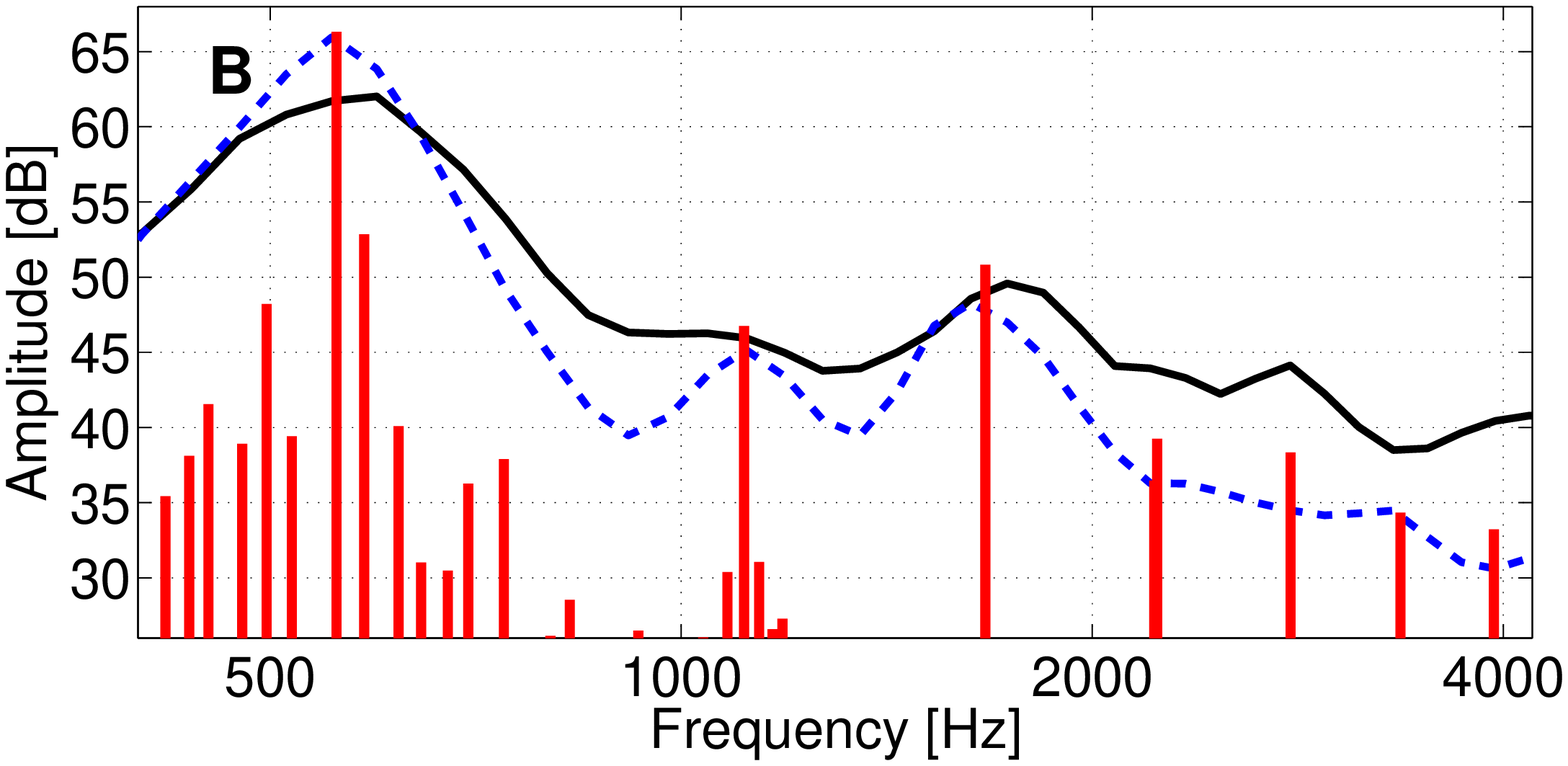}
\caption{(\textbf{A}) Low-pass filtered Hilbert envelope ($f$\textsubscript{cut-off}=$20$~Hz) for piano P1 (red line), average ICRA noise versions A (black line) and B (blue dashed line), and individual representations of the ICRA version A (gray traces). The signal--to-noise ratio for all noises was 0~dB. The envelopes were converted to dB sound pressure level (SPL). (\textbf{B}) Long-term spectra for piano sound P1 (red line) and the ICRA noise versions A (black line) and B (blue dashed line) averaged over the first 0.25~s of the waveforms.}
\label{fig:waveforms}
\end{figure}

The 3-AFC discrimination experiment described by \citet{Osses2019a} to evaluate the perceptual similarity between piano notes was simulated using the PEMO model. A general description of the experimental procedure and sound stimuli is given, indicating the adopted considerations to run our simulations. Complementary details about the experimental design can be found in our previous studies \cite{Osses2016b,Osses2017b,Osses2019a,Chaigne2019}.

\subsection{Apparatus and procedure}

The simulations were run using the AFC toolbox for MATLAB \cite{Ewert2013}. The AFC toolbox provides a framework to conduct listening experiments, including the option to enable an artificial rather than a human listener. The artificial listener consists of an auditory model (here the PEMO model) with a central processor based on signal detection theory (Sec.~\ref{sec:central-proc}).

The experiment was implemented as a 3-AFC task with the level of the ICRA noises used as adjustable parameter, expressed as signal-to-noise ratios (SNRs). The set-up of the task was almost identical to that used in the experimental sessions \cite{Osses2019a}. We only introduced small deviations to the experimental procedure, aiming at reducing the simulation time. The simulation procedure was as follows: In each adaptive track two sounds were compared, the target sound (presented once) and the reference sound (presented twice). The noise level was adjusted following a two-down one-up rule until 8 reversals were reached (4 less than in the experiments). The step sizes were set to 4~dB, 2~dB (after the second reversal) and 1~dB (after the fourth reversal). The corresponding discrimination threshold was obtained from the median of the last 4 reversals. The sound presentation level was randomly varied (roved) by uniformly-distributed levels in the range $\pm 4$~dB. The threshold estimation was repeated 6 times for each condition.

\subsection{Stimuli}
\label{sec:stimuli}


\subsubsection{Piano sounds}
\label{sec:piano-stimuli}

A selection of non-reverberant piano notes recorded from historical Viennese pianos was used for the simulations \cite[see Dataset 1,][]{Osses2019a}. In brief, recordings of note C\#$_5$ ($f_0=554$~Hz) from seven pianos were used. One recording per piano was chosen leading to a total of 7 stimuli. The waveforms had a duration of 1.3~s including a 150-ms down cosine ramp. The note onset occurred after 0.1~s reaching a maximum loudness $S$\textsubscript{max} of about 18~sone. With 7 stimuli, 21 piano pairs can be formed. For each of these 21 piano pairs, 6 thresholds were simulated, using 3 times one of the pianos as target with the other piano as reference and vice versa. 

\subsubsection{Piano-weighted noises}
\label{sec:ICRA-noises}

As in the experimental sessions, background noises that are obtained using a modified ICRA noise algorithm \cite[ICRA version A,][]{Osses2019a} were used in the simulations. The resulting ICRA noises approximately follow the spectro-temporal properties of the piano sounds, but they have a gradual spectral tilt towards high frequencies. This spectral mismatch of up to 10~dB was corrected in version B of the algorithm \cite[][as used for their Dataset~2]{Osses2019a}. This is illustrated in Fig.~\ref{fig:waveforms}. In this paper we first compared the experimental data with the simulations using ICRA noises version A. We then ran additional simulations using version B of the algorithm, where no experimental data were collected, to gain  insights into how frequency cues with similar temporal characteristics may have been integrated by the participants of the reference experiment. A detailed description of both ICRA noise algorithms can be found in our experimental paper \cite{Osses2019a}.

For the comparison between two pianos, e.g., pianos P1 and P3 (or P3 and P1) individual noises that followed the spectro-temporal properties of each piano (N1 and N3) were combined to generate a paired noise (N13). 

\subsection{Reference data: Experimental discrimination thresholds}
\label{sec:experimental-data}

Experimental thresholds thres\textsubscript{exp} using the stimuli and procedures described in this section had been previously collected from 20 participants \cite[][dataset 1]{Osses2019a}. From a total of 210 discrimination thresholds thres\textsubscript{exp}, 179 values were used (31 thresholds had been excluded after a data consistency check) to obtain the median thresholds that are indicated as red triangles in Fig.~\ref{fig:sim+exp-results}, which are shown together with their corresponding interquartile ranges (IQRs). The experimental thresholds range between thres\textsubscript{exp,max}$=20.75$~dB (pair 23) and thres\textsubscript{exp,min}$=-1.75$~dB (pair 26), with a dynamic range DR\textsubscript{exp}$=$thres\textsubscript{exp,max}$-$thres\textsubscript{exp,min}$=22.5$~dB.

\subsection{Simulations}

\subsubsection{Exploratory simulations: Subset of piano sounds}

At first, a subset of 9 (of the 21) available piano pairs was used for the simulations. These 9 pairs were chosen to be a representative sample of the range of experimental similarity between pianos, i.e., of the SNRs of thres\textsubscript{exp} (red triangles in Fig.~\ref{fig:sim+exp-results}). The selected piano pairs were: pair 12, 15, 16, 23, 26, 27, 37, 45, and 47.\footnote{Piano pairs 23 and 47 were taken from the most similar end (high SNRs at the threshold) of the similarity axis (abscissa of Fig.~\ref{fig:sim+exp-results}). Pairs 26, 27, and 37 were taken from the least similar end of the axis. The remaining pairs 12, 15, 16, and 45 were taken from the intermediate similarity range.} This reduced set of sounds was used for (1) developing our template approach  (Sec.~\ref{sec:central-proc}), and for (2) testing the duration of the ``observation (listening) period'' of the template. This latter aspect is a consequence of the lack of success (see the last column of Table~\ref{tab:res-exploratory}) to simulate the discrimination thresholds when using whole-duration piano waveforms as inputs to the model. The low thresholds in that condition were attributed to a sensitive artificial listener, who had access to more information than human listeners. As a way to remove available cues within the auditory model, the piano sounds were truncated to shorter durations. This is equivalent to reducing the observation period $t$\textsubscript{obs} of the artificial listener and can be seen as a simple way to account for a limited human-like working memory\footnote{In the artificial listener's  decision, a representation $R_x$ is correlated with a time-aligned template $T$. Any slight temporal misalignment between the two would reduce the correlation value and make the artificial observer less sensitive. One could argue that the human memory is not capable of preserving such a detailed template with a duration of 1.3~s. and that latter parts of the stored template have an increasing temporal jitter, reducing their contribution to the discrimination process. This form of information reduction is in line with the memory-decay approach adopted by \citet{Wallaert2017} who introduced a memory noise ``E\textsubscript{mem},'' that increased with the length of the internal representations. Our chosen approach of a shortened observation window should be seen as the most simple implementation of this concept.} \cite[see also,][]{Osses2018a}. 

Under the hypothesis that listeners provided a greater weighting to the piano note onset, a truncation of the piano waveforms should give a higher correlation between the simulated and experimental results. As will be shown in Sec.~\ref{sec:results}, our simulation results provide evidence to support this hypothesis.

\subsubsection{Simulations using the whole dataset of piano sounds}

The simulation of discrimination thresholds thres\textsubscript{sim} for the whole dataset of piano sounds (21 piano pairs) was run using the optimal observation period $t\textsubscript{obs}$ obtained from the exploratory simulations and the adopted template approach. These thres\textsubscript{sim} values were used to evaluate the performance of the artificial listener with respect to the reference thresholds thres\textsubscript{exp} (Section \ref{sec:experimental-data}).

\begin{table}[b]
\caption{Results of the simulations using a subset of 9 piano pairs and different $t\textsubscript{obs}$ durations. The minimum and maximum simulated thresholds are indicated together with their dynamic range (DR$=$thres\textsubscript{max}$-$thres\textsubscript{min}). The correlation between simulations and the corresponding experimental data (shown in Fig.~\ref{fig:sim+exp-results}) is given. The SNR range of the experimental data is indicated in column \textsf{Exp}.} \vspace{-12pt}
\centering
\scalebox{.95}{
\begin{tabular}{lcccccccccc} \hline\hline
			     & \multicolumn{8}{c}{``Observation (listening) period'' $t\textsubscript{obs}$ (s)} \\
Parameter                    & Exp. & 0.2   & 0.25 & 0.3   & 0.5  & 0.7  & 0.9  & 1.3  \\ \hline
thres\textsubscript{max} (dB)& 20.75 & 15.0  & 20.5 & 14.25 & 9.25 & 5.0  & 3.25 & 2.0 \\
thres\textsubscript{min} (dB)& -1.75 & -0.25 &-1.0  & 1.5   &-0.5  &-1.25 &-1.75 &-3.0 \\
DR                       (dB)& 22.5  & 15.25 & 21.5 & 12.75 & 9.75 & 6.25 & 5.0  & 5.0 \\
Pearson $r_p$  & $-$   & 0.66\footnotemark[1] & 0.71\footnotemark[1]& 0.65\footnotemark[2]& 0.34 & 0.45 & 0.25 & -0.21\\  
Spearman $r_s$  & $-$   & 0.60\footnotemark[2]& 0.78\footnotemark[1]& 0.47  & 0.11 & 0.49 & 0.21 & 0.09\\ \hline\hline
\end{tabular}
}
\footnotetext[1]{Significant correlation, $p<0.05$, $N=9$.} 
\footnotetext[2]{Correlations that approach significance, $p<0.10$, $N=9$.}
\label{tab:res-exploratory}
\end{table}

\subsubsection{Extra simulations using ICRA noises version B}

Simulations using ICRA noises version B were run for the whole dataset of pianos using the optimal observation duration $t\textsubscript{obs}$. This choice allows to quantify the perceptual difference between ICRA noise versions, illustrated in Fig.~\ref{fig:waveforms}\textbf{B}, that we hypothesized to be small based on our previous comparison between non-reverberant (using version A) and reverberant piano sounds (using version B) \cite{Osses2019a}.

\section{Results}
\label{sec:results}



\begin{figure}
	\centering
	\includegraphics[width=0.48\textwidth]{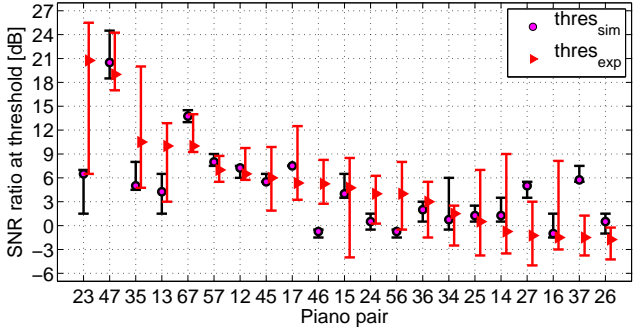}
	\caption{Discrimination thresholds using the whole dataset of piano sounds (21 piano pairs). The median simulated thresholds thres\textsubscript{sim} are indicated by the magenta circle markers. The red triangle markers correspond to the experimental thresholds thres\textsubscript{exp} (see Sec.~\ref{sec:experimental-data}). The thresholds are shown together with their IQRs. The piano pairs along the abscissa are ordered from higher to lower SNR thresholds based on the experimental data.}
	\label{fig:sim+exp-results}
\end{figure}

\subsection{Exploratory simulations: Subset of piano sounds}

The simulation results for the selection of 9 piano pairs are shown in Table~\ref{tab:res-exploratory}. In the table, information about the minimum (lowest median) and maximum (highest median) estimated thresholds is shown and their difference is indicated as a dynamic range (DR) in dB.  The simulations with $t\textsubscript{obs}$=1.3~s delivered thresholds between thres\textsubscript{sim,max}$=2$~dB and thres\textsubscript{sim,min}$=-3$~dB with a DR of $5$~dB. Such low thresholds with respect to the thres\textsubscript{exp} values indicate that the artificial listener had access to more information than the actual participants with $t\textsubscript{obs}$=1.3~s. As a way to remove available cues within the model, the observation period $t\textsubscript{obs}$ of the artificial listener was limited to shorter durations ($t\textsubscript{obs}$ of 0.20, 0.25, 0.3, 0.5, 0.7, 0.9~s), omitting the latter tails of the representations $R_x$ from the CCV assessment (Eq.~\ref{eq:CCV-with-temp}). The results for $t\textsubscript{obs}$=0.9 and 1.3~s had a constant DR of 5~dB, and for shorter durations, thres\textsubscript{sim,max} increased  reaching a maximum DR of 20.5~dB for $t\textsubscript{obs}$=0.25~s. For the shortest tested duration of 0.20~s the DR decreased by 6~dB. The interpretation of these results is that at $t\textsubscript{obs}$=0.25~s the piano sounds were judged by the artificial listener as most distinct. Given that this duration $t\textsubscript{obs}$ also had the best fit with the experimental data (Pearson correlation $r_p$=0.71, $p$=0.03; Spearman correlation $r_s$=0.78, $p$=0.02, for $N$=9), $t\textsubscript{obs}$=0.25~s was further used to simulate the discrimination thresholds of the remaining 13 piano pairs.


\subsection{Simulations using the whole dataset of piano sounds}
\label{sec:sim-whole}

The discrimination thresholds using the whole dataset of piano sounds (21 piano pairs) were simulated using the first $t\textsubscript{obs}$=0.25~s of waveforms, based on the results of the exploratory simulations. The median thresholds thres\textsubscript{sim} are indicated as magenta circle markers in Fig.~\ref{fig:sim+exp-results}. The thresholds are shown together with their interquartile ranges (IQRs). The simulations at this duration ($t\textsubscript{obs}$=0.25~s) were not only highly correlated with the experimental data but they also had a comparable DR=$21.5$~dB (same DR as in the exploratory analysis). The thres\textsubscript{sim} values range between thres\textsubscript{sim,max}$=20.5$~dB (pair 47) and thres\textsubscript{sim,min}$=-1$~dB (pair 16). The discrimination thresholds thres\textsubscript{sim} and thres\textsubscript{exp} were significantly correlated with a Spearman (rank-order) correlation $r_s$=0.63, $p<$0.001, $N$=21, and a Pearson correlation $r_p$=0.54, $p$=0.02, $N$=19.\footnote{For the assessment of the Pearson correlation, the data of two piano pairs (pairs 23 and 47) had to be excluded to meet the assumption of data normality. The two excluded pairs had both SNR thresholds above 12~dB.} 

\subsection{Extra simulations using ICRA noises version B}
\label{sec:res-ICRA-A-B}

\begin{figure}
	\centering
	\includegraphics[width=0.48\textwidth]{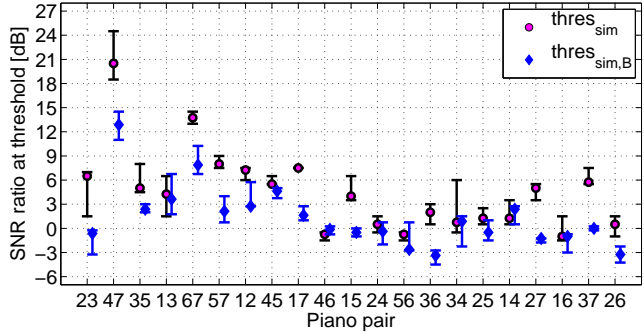}
	\caption{Simulated thresholds using the whole dataset of piano sounds (t\textsubscript{obs}=0.25~s) and different types of ICRA noise. The median simulated thresholds thres\textsubscript{sim} using noises version A are indicated by the magenta circle markers (as in Fig.~\ref{fig:sim+exp-results}). The blue diamond markers correspond to simulated thresholds thres\textsubscript{sim,B} obtained using noises version B. The piano pairs along the abscissa are ordered as in Fig.~\ref{fig:sim+exp-results}.} %
	\label{fig:sim-ICRA-A+B}
\end{figure}

\begin{figure*}
	\centering
\includegraphics[height=0.18\textheight]{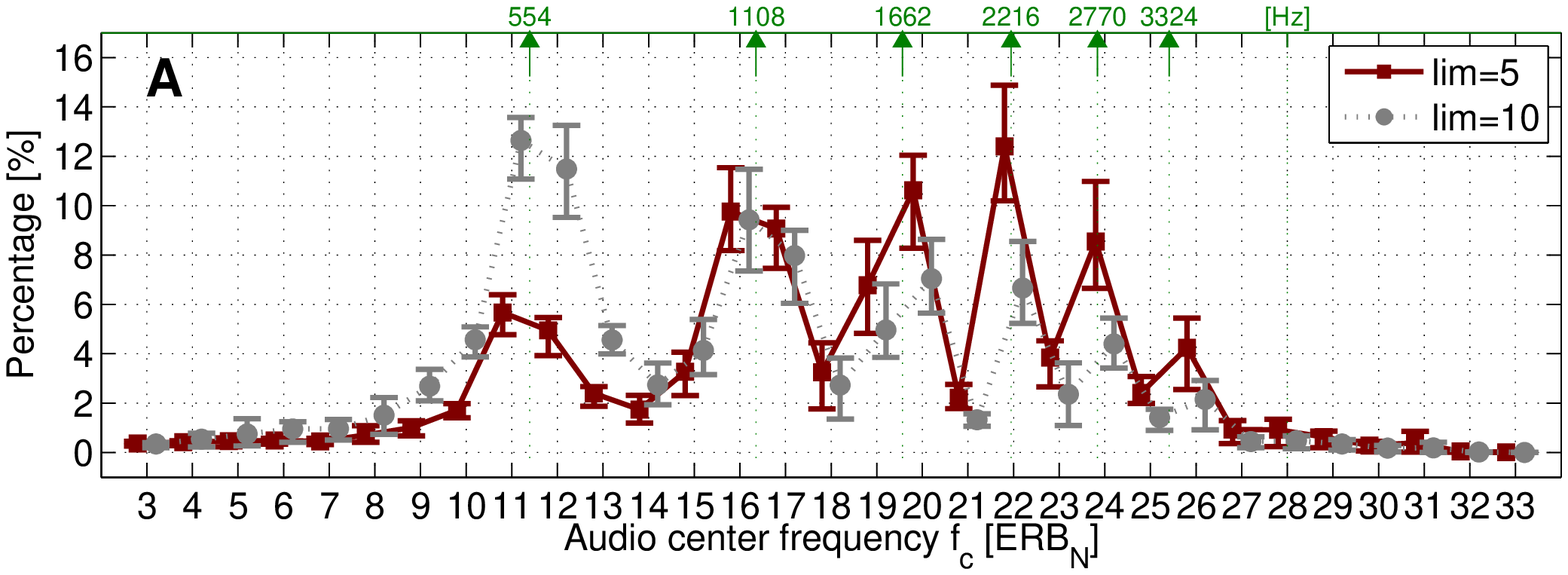}
\includegraphics[height=0.18\textheight,clip=true,trim=.75cm 0 0 0]{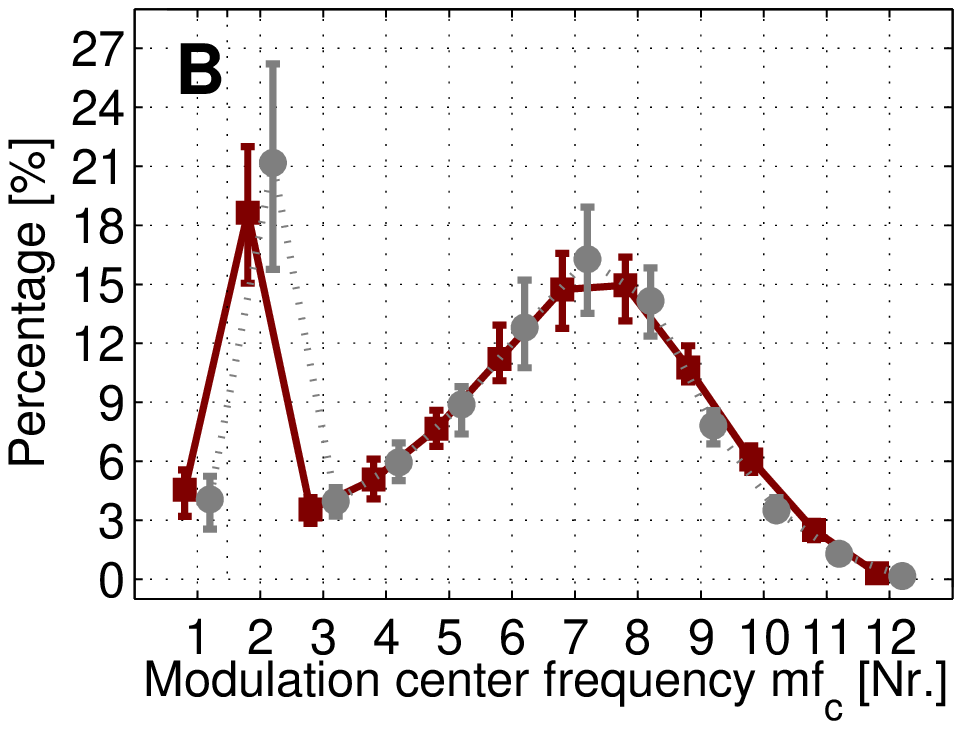}
	\caption{Average weighting of information in difference representations ($\Delta R_x \cdot T_p$) for limiter factors of lim=5 (maroon square markers) and \textsf{lim}=10 (gray circle markers) using $t\textsubscript{obs}$=0.25~s. The weighting of information in each audio ($I_m/I\textsubscript{tot}$) and modulation frequency band ($I_k/I\textsubscript{tot}$) is shown in panels \textbf{A} and \textbf{B}, respectively. The values per band are expressed as percentages. The points along the ERB scale that correspond to $f_0=554$~Hz and its first harmonics are indicated by the green numbers along the top axis.}
	\label{fig:weight-per-band-5-10}
\end{figure*}

The discrimination thresholds thres\textsubscript{sim,B} using the whole dataset of piano sounds were simulated using $t\textsubscript{obs}$=0.25~s and ICRA noises version~B, following the same simulation procedure as previously described. The median thresholds thres\textsubscript{sim,B} are indicated as blue diamond markers in Fig.~\ref{fig:sim-ICRA-A+B}. For ease of comparison, the thresholds obtained using noises version A (thres\textsubscript{sim}) were replotted from Fig.~\ref{fig:sim+exp-results}. All thresholds are shown together with the corresponding IQRs derived from 6 simulation runs. The thres\textsubscript{sim,B} values ranged between thres\textsubscript{sim,B,max}$=12.9$~dB (pair 47) and thres\textsubscript{sim,B,min}$=-3.4$~dB (pair 36). The thresholds thres\textsubscript{sim} and thres\textsubscript{sim,B} were significantly correlated with a Spearman correlation $r_s$=0.62, $p$=0.003, $N$=21, and a Pearson correlation $r_p$=0.52, $p$=0.02, $N$=19.

\section{Data analysis and discussion}
\label{sec:discussion}

The simulated thresholds thres\textsubscript{sim} of the instrument-in-noise test were significantly correlated with the experimental thresholds thres\textsubscript{exp} when only the initial part of the waveforms was used. Two aspects that affected the internal representation of the sounds leading to the obtained thres\textsubscript{sim} values are addressed in this section: (1) The weighting of information in each (audio and modulation) frequency band, and; (2) the concept of ``optimal detector'' used in the central processor stage and how its performance was affected by shortening the duration of the observation period and by the sources of variability in the model (internal) and in the stimuli (external). Finally, the effect of using ICRA noises  A and B on the simulated discrimination thresholds is discussed in terms of threshold shifts for the whole set of piano sounds.

\subsection{Information-based analysis of internal representations} 

\begin{figure}
	\centering
\includegraphics[width=0.4\textwidth]{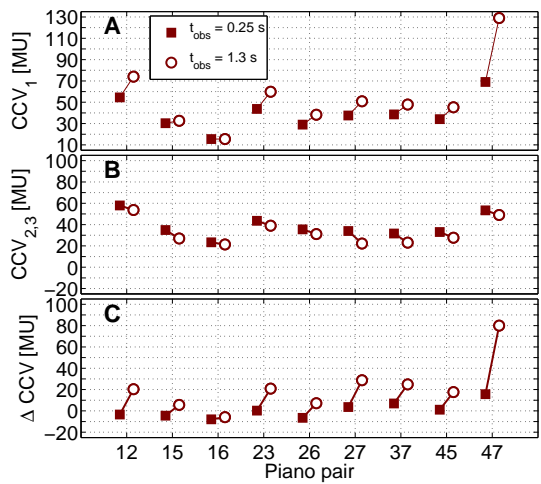}	
	\caption{CCV values for each piano pair (SNR at threshold) using $t\textsubscript{obs}$=0.25~s (filled squares) and $t\textsubscript{obs}$=1.3~s (open circles) of the internal representations. (\textbf{A}) CCV values for the target interval ($x$=1) of the leading criterion (CCV$_t$ or CCV$_r$, Eq.~\ref{eq:CCV-criterion}). (\textbf{B}) CCV values for the reference interval ($x$=2, 3) of the same criterion. (\textbf{C}) Difference between CCV values.}
	\label{fig:CCV-lead}
\end{figure}


The weighting of information for each audio and modulation frequency band within the PEMO model is primarily introduced by the dot product between internal representations and templates ($T_{p,t}$ and $T_{p,r}$, Eq.~\ref{eq:CCV-with-temp}). Following a similar approach to that used to analyze the representation of piano P1 (Fig.~\ref{fig:energy-per-band}), the contribution of each frequency band ($I_m/I\textsubscript{tot}$ and $I_k/I\textsubscript{tot}$) can be assessed using Eqs.~\ref{eq:Ik} and \ref{eq:Itot}. The following conditions were considered here: (1) when the adaptation loops are limited using a factor lim=$5$ (as suggested in this study) and with lim=$10$, and (2) considering the total duration of the piano-plus-noise sounds (1.3~s) and when only the first $t\textsubscript{obs}$=0.25~s are evaluated. In this analysis, all 21 piano pairs were included. Since our interest is on the weighting of information at threshold, the difference representation $\Delta R=R-R_N$ was assessed at the ICRA noise level indicated by thres\textsubscript{sim}. 

The information-weighted values ($I_m/I\textsubscript{tot}$ and $I_k/I\textsubscript{tot}$) for the comparison between limiter factors are shown in Fig.~\ref{fig:weight-per-band-5-10}. The values $I_m$ and $I_k$ were obtained as the median of 42 values (21 pairs with one value using $T_{p,t}$ and one value using $T_{p,r}$). The error bars indicate IQRs. The weighting $I_m/I\textsubscript{tot}$ shown in Fig.~\ref{fig:weight-per-band-5-10}\textbf{A} shows that using a stronger limiter factor (here lim=5), the information of higher audio frequency bands receive a higher relative weighting. For lim=10, the weighting seems to be very similar to the distribution of information for the piano-alone representation shown in Fig.~\ref{fig:energy-per-band}\textbf{A}. 

The information contribution of each modulation band is shown in Fig.~\ref{fig:weight-per-band-5-10}\textbf{B}. The second modulation band (mf$_c$=$5$~Hz) had a weighting of 18.6\% for the representations with lim=$5$, which is 2.6\% below the weighting for the same band when lim=10 was used (weighting of 21.2\%). The first modulation band had a low weighting despite its high value in the piano-alone representation (Fig.~\ref{fig:energy-per-band}\textbf{B}). This result indicates that the slow envelope changes tracked in this modulation band did not differ considerably from piano to piano. Bands 6 to 9 showed a slight increase in their weighting for lim=$5$ (compared to the lim=$10$), while the rest of the bands had a similar weighting with both limiter factors. 

The band weighting values for the comparison between signal durations ($t\textsubscript{obs}$=$0.25$~s and $1.3$~s) were very similar (mean difference $\Delta I_m/I\textsubscript{tot}$ of 0.00\%, IQR of 0.32\%, not shown here) and, therefore, they seemed to be unaffected by the duration $t\textsubscript{obs}$ of the piano sounds. To explain the influence of $t\textsubscript{obs}$ on the simulated thresholds, the performance of the artificial listener was analyzed in terms of other factors, as detailed next. 

\subsection{Factors that affected the simulated performance}

\subsubsection{Reducing the performance of the optimal detector}

The central processor of the PEMO model is inspired by the concept of an optimal detector. In signal detection theory, the term optimal refers to the fact that the detector has the best possible performance given specific stimulus properties. In other words, if a cue is available in the stimulus, then the detector uses it \cite[][Ch.~6]{Green1966}. For this reason, detectors that are optimal can be used as baselines for human detection. The results of our exploratory simulations  showed that the participants' performance in the instrument-in-noise experiment is below the ideal performance, where simulated thresholds for whole-duration sounds covered a range of only $5$ dB (Table~\ref{tab:res-exploratory}). 

One way to bring the simulated thresholds to a range closer to that of the experimental data is the removal of ``evidence'' from the stimuli, which is assumed to be cumulative during the observation period $t\textsubscript{obs}$ of the artificial listener. The simulated thresholds for shorter $t\textsubscript{obs}$ durations resulted in thresholds with a higher dynamic range (DR=thres\textsubscript{max}$-$tres\textsubscript{min}), increasing from $5$~dB for $t\textsubscript{obs}$=$1.3$~s to $21.5$~dB for $t\textsubscript{obs}$=$0.25$~s. To evaluate this DR increase, an analysis based on CCV values is presented using the $t\textsubscript{obs}$ durations of 0.25~s and 1.3~s.

The CCV values expressed in model units (MU) for the subset of 9 piano pairs are shown in Fig.~\ref{fig:CCV-lead} at a noise level given by their corresponding thres\textsubscript{sim} value, with filled and open markers indicating CCV values using $t\textsubscript{obs}$ durations of 0.25~s or 1.3~s, respectively. For this analysis no level roving was applied, meaning that at the exact CCV values of the figure the artificial listener would obtain discrimination scores of 70.7\% or slightly above.\footnote{This is due to the overall lower thresholds (i.e., better discriminability) when removing the level roving, as can be seen in \textsf{no-rove} thresholds of Fig.~\ref{fig:int-ext-noise}.} In general, at these noise levels only one of the two decision criteria of Eq.~\ref{eq:CCV-criterion} failed. The criterion that failed first was labeled as ``leading criterion.'' The CCV values of the leading criterion for target and reference intervals are shown in Fig.~\ref{fig:CCV-lead}\textbf{A} and Fig.~\ref{fig:CCV-lead}\textbf{B}, respectively, and their difference $\Delta$CCV is shown in Fig.~\ref{fig:CCV-lead}\textbf{C}. The $\Delta$CCV values ranged between $-7.9$ MU (pair 16) and $15.7$ MU (pair 47) for representations with $t\textsubscript{obs}$=0.25~s and between $-5.9$ MU (pair 16) and $80$ MU (pair 47) for representations with $t\textsubscript{obs}$=1.30~s. These $\Delta$CCV values indicate that the discriminability between pianos either remained approximately unchanged (pair 16) or improved with $t\textsubscript{obs}$ (pairs 12, 15, 23, 26, 27, 37, 45, and 47) and that the use of shorter internal representations compressed the $\Delta$CCV\textsubscript{0.25~s} values without changing significantly the relative discriminability between pianos, having a rank-order correlation of $r_s$=$0.92$, $p$=$0.001$, $N$=$9$ with respect to the $\Delta$CCV\textsubscript{1.30~s} values. The differences $\Delta$CCV\textsubscript{0.25~s} were, however, susceptible to the variance introduced by the internal noise. Since each CCV value was varied by a number drawn from a normal distribution with standard deviation $\sigma$ of $10.1$~MU, the difference $\Delta$CCV values were also normally distributed, but with a standard deviation of $\sqrt{\sigma^2+\sigma^2}=14.4$~MU. Eight of the 9 difference $\Delta$CCV\textsubscript{0.25~s} values in Fig.~\ref{fig:CCV-lead}\textbf{C} (20 of 21 if the whole dataset were used) lie in the variability range of the internal noise ($\pm 14.4$~MU). This means that the internal noise played a prominent role in the discrimination performance of the artificial listener. For representations with $t\textsubscript{obs}$=1.3~s a much larger variance of the internal noise would be needed for reaching thresholds in a similar SNR range. Although it was possible to introduce a higher variability to the internal representations, this would have strongly limited the performance of the PEMO model, reducing its predictive power for other already validated auditory tasks \cite[e.g.,][his App. D]{Dau1997b,Osses2018}, violating the so-called backward compatibility of the PEMO model.

\subsection{Effect of the sources of variability} 
\begin{figure}
	\centering
	\includegraphics[width=.42\textwidth]{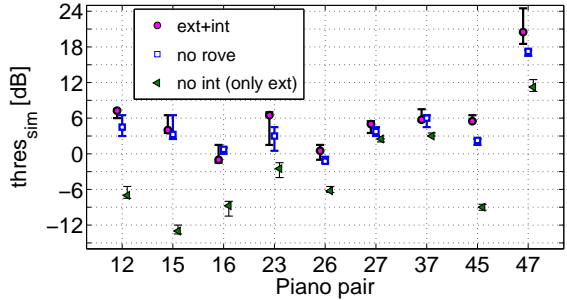}
	\caption{Simulated thresholds thres\textsubscript{sim} for the subset of 9 piano pairs in the following conditions: (1) considering internal and external sources of variability (magenta circle markers, as in Fig.~\ref{fig:sim+exp-results}); (2) with internal variability but without level roving (square blue markers); (3) without internal variability, i.e., considering only sources of external variability.}
	\label{fig:int-ext-noise}
\end{figure}

In order to quantify the influence of the sources of variability on the obtained thresholds thres\textsubscript{sim}, simulations for the subset of 9 piano pairs using $t\textsubscript{obs}$=0.25~s were run in the following conditions: (1) No level roving (\textsf{no--rove} condition), i.e., using only the internal noise variability and running noises, and (2) No internal noise (\textsf{no--int} condition), i.e., using only sources of external variability (level rove and running noise). The resulting median thresholds (of 6 estimates) with their IQRs are indicated by the blue squares and the green triangles in Fig.~\ref{fig:int-ext-noise}, respectively. The simulated thresholds using both sources of variability (as shown in Fig.~\ref{fig:sim+exp-results}) are indicated by the magenta circle markers (\textsf{ext+int} condition) and were used as a baseline for this analysis. The simulated thresholds in the \textsf{no--rove} condition followed the same trend as the \textsf{ext+int}-thresholds (correlation of $r_s$=0.77, $p$=0.02, $N$=9) and differed by 3.5~dB (pair 23) or less. This was not the case for the thresholds in the \textsf{no--int} condition, that were much lower than the \textsf{ext+int}-thresholds and were not significantly correlated ($r_s$=$0.53$, $p$=$0.15$, $N$=$9$). This means that the limit in performance introduced by the sources of external variability of the instrument-in-noise task was not sufficient to explain the performance of the artificial listener. This analysis provides further evidence of the dominant role played by the internal noise in the decision of the artificial listener for representations with $t\textsubscript{obs}$=0.25 s.

%
%
%

\subsection{Simulated thresholds for different ICRA noise versions}
The simulated thresholds thres\textsubscript{sim,B} of the instrument-in-noise method using ICRA noises version B (Fig.~\ref{fig:sim-ICRA-A+B}) were significantly correlated with the thres\textsubscript{sim} values obtained using noises version~A ($r_s$=$0.62$, and $r_p$=$0.52$, see Sec.~\ref{sec:res-ICRA-A-B}). The difference between simulated thresholds ($\Delta$SNR=thres\textsubscript{sim}$-$thres\textsubscript{sim,B}) is shown in Fig.~\ref{fig:diff-ICRAs}. The median difference across all piano pairs was $3.75$~dB (horizontal gray dashed-dotted line in the figure) with an IQR between $0.8$ and $5.9$~dB. This means that on average, noises version A produced higher discrimination thresholds (thres\textsubscript{sim}$>$thres\textsubscript{sim,B}). 

To investigate how the two ICRA noise types influenced the threshold estimations, we classified the piano pairs into three groups, defined by the shadowed area in Fig.~\ref{fig:diff-ICRAs}: (1) Pairs with $\Delta$SNR values that were equal to or greater than percentile 75 ($\Delta$SNR$\geq5.9$~dB): pairs 17, 23, 27, 47, 57, 67; (2) Pairs with differences within the IQR: pairs 12, 15, 24, 25, 26, 35, 36, 37, 45, 56; and (3) Pairs with differences that were equal to or lower than percentile 25 ($\Delta$SNR$\leq0.8$~dB): pairs 13, 14, 16, 34, and 46. One piano pair of each group was further analyzed: pair 47 ($\Delta$SNR$=7.62$~dB), pair 26 ($\Delta$SNR$=3.75$~dB), and pair 14 ($\Delta$SNR$=-1.13$~dB). 

The spectrum of the noises related to the selected pairs (N47, N26, N14) expressed as band levels (BL) within 1~ERB are shown in Fig.~\ref{fig:diff-BL}. This figure shows that the noise spectra (ICRA version A and B) related to paired noise N47 (Group~1) have similar BLs at around $4 f_0$ (Fig.~\ref{fig:diff-BL}\textbf{A}), and that for noise N14 (Group 3) this is the case at around $f_0$ (Fig.~\ref{fig:diff-BL}\textbf{C}). The spectra related to noise N26 (Group 2) have similar BLs between $f_0$ and $4 f_0$ (between 13.4 and 21.4 ERB$_N$ in Fig.~\ref{fig:diff-BL}\textbf{B}). We may therefore infer that the frequency region where noises A and B produce a similar masking, i.e., when $\Delta$BL$\approx$0 (see the open makers in Fig.~\ref{fig:diff-BL}\textbf{D}), is related to the most important (audio) frequency range during the simulations and that a similar frequency weighting would have been used by the participants during the experimental sessions. This analysis is based on the fact that:

\begin{figure} 
	\centering
	\includegraphics[width=0.46\textwidth]{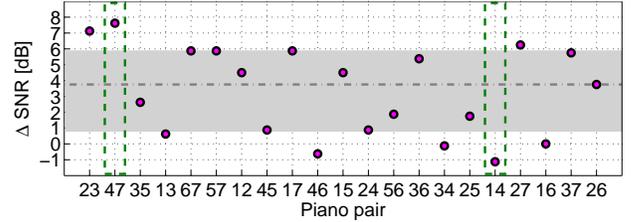}
	\caption{Difference $\Delta$SNR between simulated thresholds thres\textsubscript{sim} and thres\textsubscript{sim,B} (Fig.~\ref{fig:sim-ICRA-A+B}) that were obtained using ICRA noises version A and B, respectively. A $\Delta$SNR value above 0~dB indicates that the SNR threshold thres\textsubscript{sim} is higher than the thres\textsubscript{sim,B} for the corresponding piano pair. The shadowed area indicates the IQR of the difference that had a median of $3.75$~dB (gray dashed-dotted line).} 
	\label{fig:diff-ICRAs}
\end{figure}
%

\begin{itemize}[wide, labelwidth=!, labelindent=0pt]
\item Both ICRA noise versions have the same statistical temporal properties (Fig.~\ref{fig:waveforms}\textbf{A}) and differ only in their weighting towards high auditory bands in version~A ($+10$~dB in the highest band with respect to the $f_0$-centered band, Fig.~\ref{fig:waveforms}\textbf{B}). This means that the expected masking difference when using ICRA noises A and B should be at most $10$~dB, which seemed to be the case (maximum difference of $7.62$~dB for pair 47, Fig.~\ref{fig:diff-ICRAs}).
\item The masking efficiency of both ICRA noises has been experimentally validated for ``Dataset~1'' (same stimuli as used in this study) and Dataset 2 (reverberant piano sounds) for version A and version B noises, respectively \cite{Osses2019a}. Moreover, this paper provides simulations using noises version A, whereas simulations using noises version B were published by \citet{Osses2018a} using the PEMO model with the exact same configuration as presented in this paper. The correlation between experimental data and simulations were similar (Dataset 1 with simulations shown in Sec.~\ref{sec:sim-whole}: $r_s$=0.63, $p<$0.001, $N$=21; Dataset~2 by \citealt{Osses2018a}: $r_s$=0.61, $p<$0.001, $N$=21) suggesting that the PEMO model performance is the same with either ICRA noise version and, hence, the model is able to follow overall changes in measured psychoacoustic performance.
\end{itemize}

\begin{figure}
\includegraphics[width=0.46\textwidth,trim=0 1.3cm 0 0,clip=true]{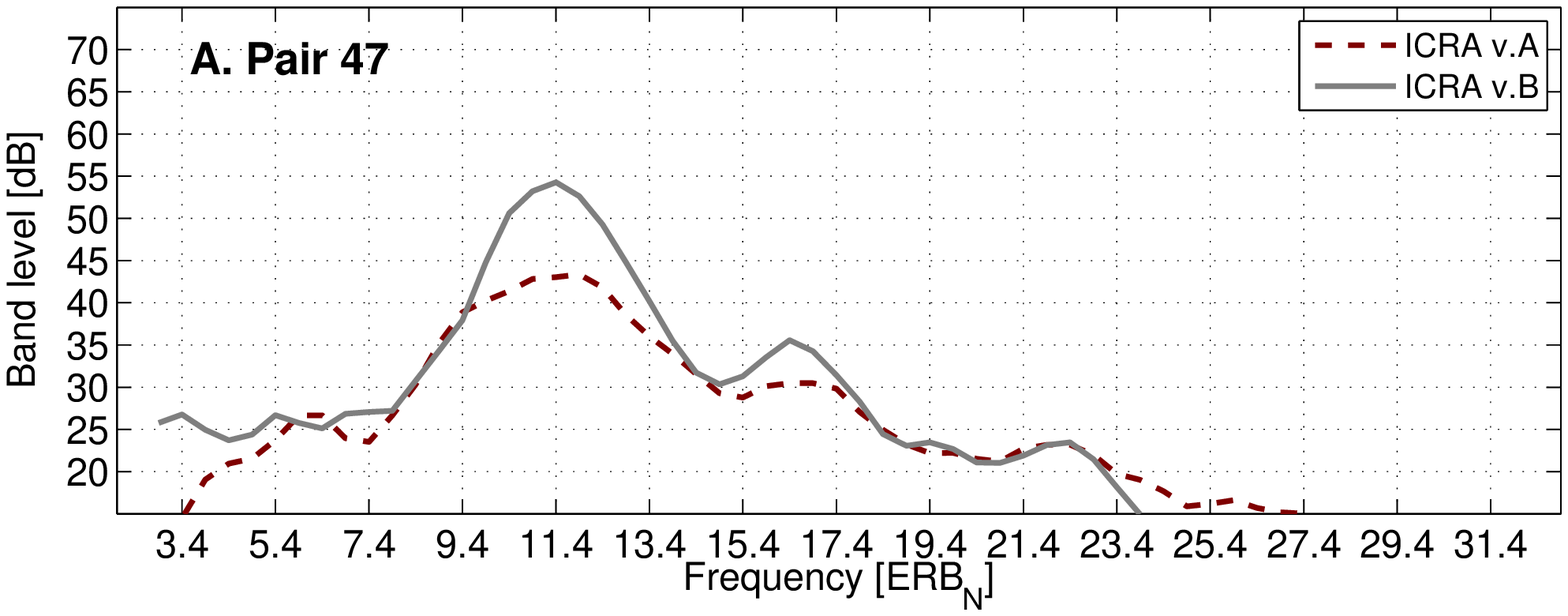}
\includegraphics[width=0.46\textwidth,trim=0 1.3cm 0 0,clip=true]{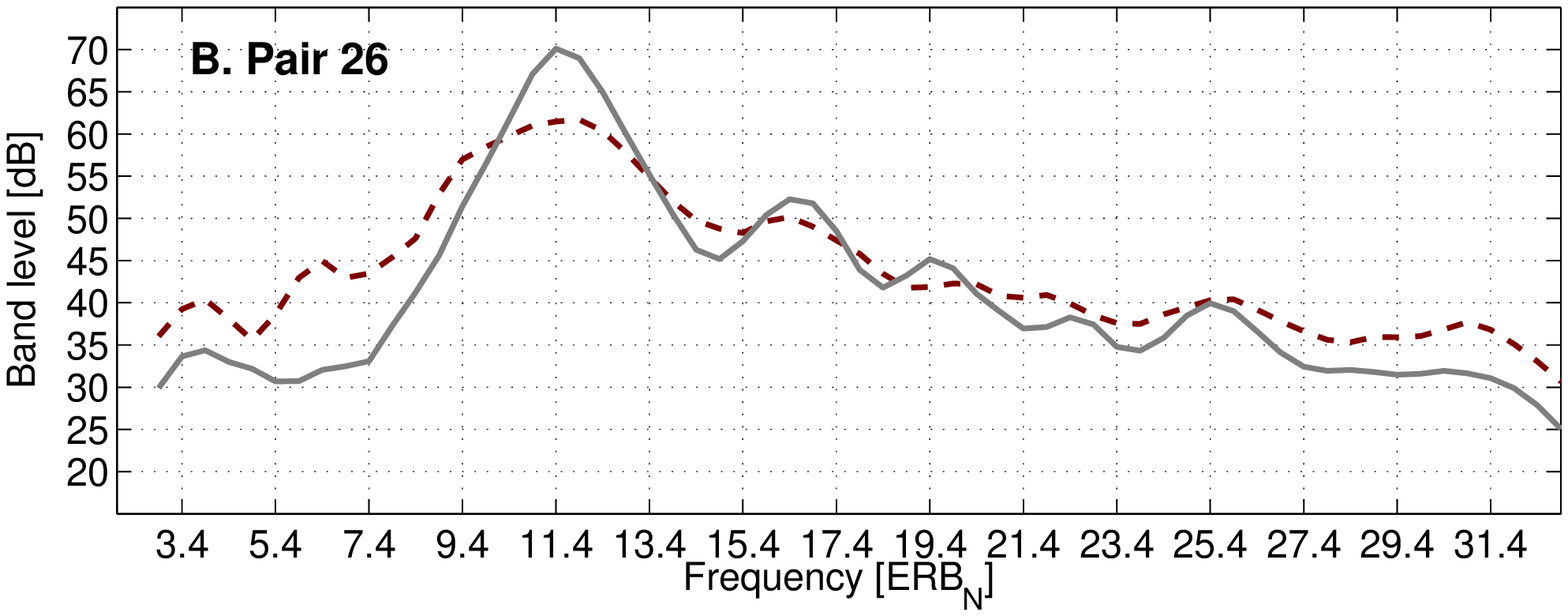}
\includegraphics[width=0.46\textwidth,trim=0 0 0 0,clip=true]{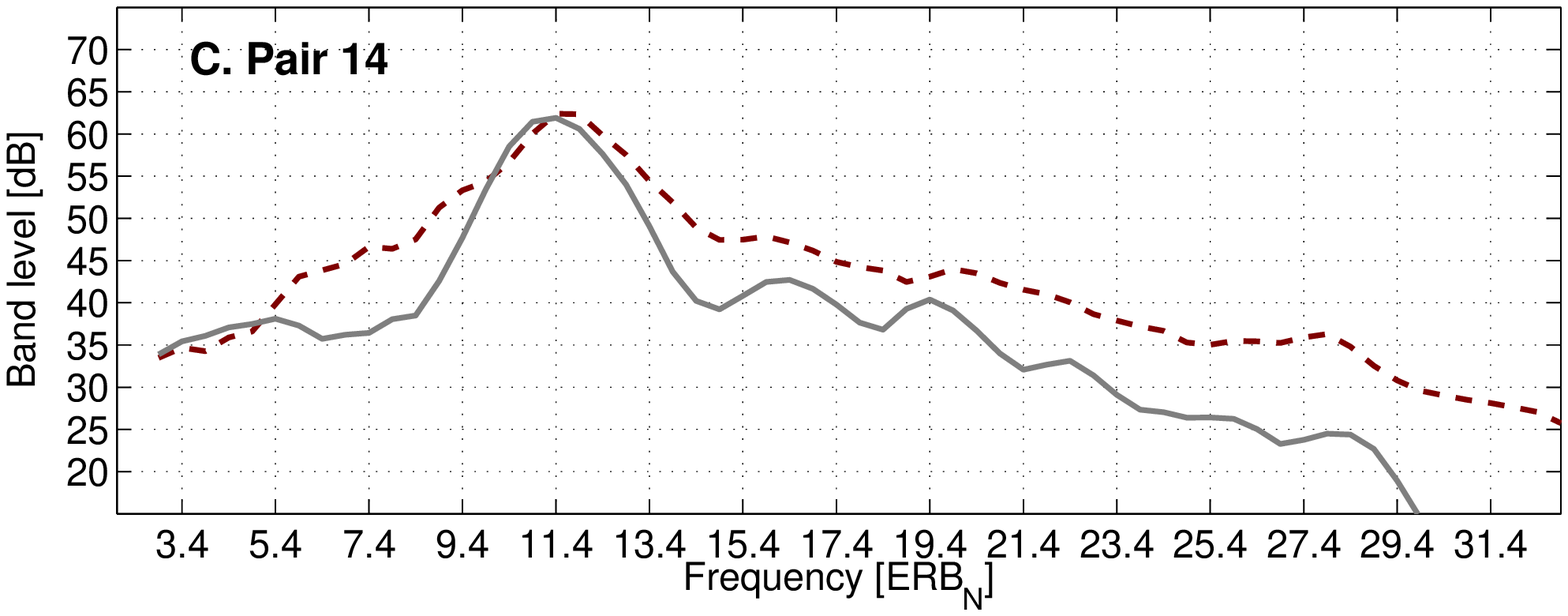}
\includegraphics[width=0.46\textwidth]{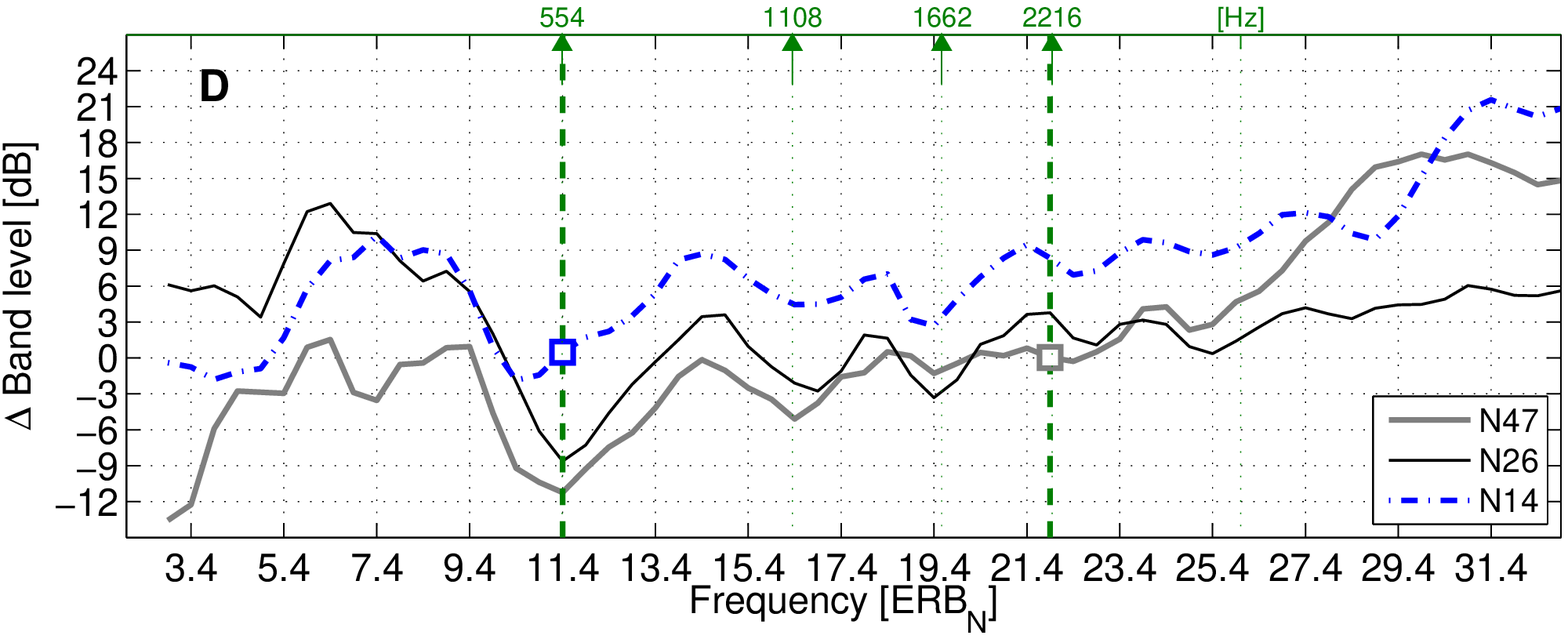}
	\caption{(\textbf{A,B,C}) Band levels (BL) for the paired noises N47, N26, and N16, generated using ICRA noises version A (maroon dashed line) and B (gray solid line) at the corresponding simulated threshold, with $t\textsubscript{obs}=$0.25~s. (\textbf{D}) Difference between band levels (BL\textsubscript{ICRA v.A}$-$BL\textsubscript{ICRA v.B}) of the paired noises N47 (gray thick solid line), N26 (black thin line), and N14 (blue dashed-dotted line), respectively. The masking of the ICRA noises A and B is approximately the same ($\Delta$BL$\approx 0$~dB) at frequencies of $4\cdot f_0$ and $f_0$ for noises N47 and N14 (open markers), respectively. For noise N26 the masking of both noises is relatively similar between frequencies of 13.4 and 21.4~ERB$_N$ with $\Delta$BL values close to 0~dB ($\pm 3$ dB), a region that is roughly between $f_0$ and $4\cdot f_0$.} 
	\label{fig:diff-BL}
\end{figure}

As for further evaluations using ICRA noises (Sec. \ref{sec:ICRA-noises}), both noise versions A and B were able to efficiently mask the spectro-temporal properties of the test piano sounds. Yet, if the  noise efficiency is to be evaluated in terms of the amount of noise needed to mask the properties of the piano sounds, then noises version~A perform better because for the same overall (broad-band) noise level the discrimination thresholds have on average higher SNRs (lower noise level) compared to noises version B: $\Delta$SNR$=3.75$~dB (Fig.~\ref{fig:diff-ICRAs}). This ``better performance'' is, however, at the expense of a gradual level mismatch towards higher frequencies of the noises with respect to the sounds to be masked. Therefore, if the efficiency of the noises is to be evaluated in terms of how well do the spectro-temporal properties of the noise follow the properties of the sounds to be masked, then noises version B perform better. 

\section{Conclusions}

In this study a long-tradition model of the auditory periphery, the perception model (PEMO) \cite{Dau1997b}, was used to simulate the perceptual similarity between recorded sounds of one note (C\#$_5$) played on 7 different pianos. Each stage of the model was described indicating the set of configuration parameters we used. The use of sounds with strong onset characteristics required us to include an in-depth analysis of the auditory adaptation stage, the adaptation loops, whose properties are often described at a high level and the details of their implementation and properties are scarce. In this paper we showed that the overshoot limitation factor is not directly related to the ratio between onset and steady-state responses of the adaptation loops as typically claimed in the literature (Appendix~\ref{app:adaptation}), and that the use of a limiter factor of 5 (instead of 10) produces an onset to steady-state ratio $\Psi\textsubscript{onset}/\Psi\textsubscript{steady}$ that is closer to the physiological observations by \citet{Westerman1984} that \citet{Muenkner1993} used as reference to implement the overshoot limitation of the adaptation loops.

The simulated task for the piano comparisons was a 3-AFC discriminability task conducted in a background noise, whose experimental (reference) data were recently reported by \citet{Osses2019a}. To simulate how similar two piano sounds were, the back-end stage of the model required the use of two memory templates. The simulated discriminability thresholds thres\textsubscript{sim}, expressed as signal-to-noise ratios, were significantly correlated with the reference thresholds thres\textsubscript{exp}, but this required a considerable reduction of accessible information to the artificial observer. This information reduction was achieved by limiting the observation period of the artificial observer to the sound onsets, with an optimal duration $t\textsubscript{obs}$ of 0.25~s. Such an approach can be interpreted as an attentional trigger that assumes non-useful information in the piano internal representations for $t>t\textsubscript{obs}$. This approach is comparable to the (more elaborate) memory-noise concept employed by \citet{Wallaert2017}. Further analyses were presented to quantify the influence of an additive internal noise, which dominated the model performance for the obtained $t\textsubscript{obs}$, and how other sources of external variability (level roving and running noises), to a lesser extent, limited the model performance in the instrument-in-noise task.

Finally, an information-based analysis at simulated discriminability thresholds using an alternative  noise (ICRA version B), with similar temporal but slightly different spectral properties, was used to show how the piano spectra were likely weighted during the (reference) experimental sessions. With this analysis we identified that the spectral information between the fundamental frequency of the note $f_0$=554~Hz and the partial at $4 f_0$ was most relevant to resolve the piano comparisons.

The results presented in this paper support the idea that the unified framework offered by the PEMO model can be used to evaluate perceptual tasks using complex sounds. This can be seen as an extension of the use of this type of models and their success relies on the adjustment of the central processor stage included within the model, in combination with an appropriate representation of sources of internal noise.

\begin{acknowledgments}

This research work was funded by the European Commission (EC) within the ITN Marie Sk\l{}odowska-Curie Action project BATWOMAN under the Seventh Framework Programme (EC Grant No.~605867).

\end{acknowledgments}

\appendix

\section{PEMO model in the AMT toolbox}
\label{app:PEMO-in-AMT}

\begin{table*}
\caption{Parameters of the PEMO model [\citet{Dau1997b}, \citet{Verhey1999}, `Current'],  or models that were developed based on it [\citet{Breebaart2001a}, \citet{Jepsen2008}], as implemented in the AMT toolbox (as of v0.10). The models are labeled by the last name of the first author followed by the year of publication of the corresponding reference. The nomenclature for specific sets of stage parameters within AMT are indicated in quotation marks.} \vspace{-12pt}
\scalebox{0.9}{
\begin{tabular}{llccccccc} \hline\hline
\multirow{2}{*}{ } & \multirow{2}{*}{Stage / Description} & \multicolumn{4}{c}{Model} \\ 
      &              & Current & Dau (1997) & Verhey (1999) & Breebaart (2001) & Jepsen (2008) \\ \hline    
\multicolumn{2}{l}{AMT function (*\_preproc.m)} & dau1997 & dau1997 & dau1997 & breebaart2001 & jepsen2008 \\ \hline
1-2 /& Outer, middle ear, cochlear filters & `gtf\_osses2020' & `gtb\_dau'\footnotemark[1] & `gtb\_dau'\footnotemark[1] & default\footnotemark[2] & default \\
  & Outer, middle ear & yes & No       & No         & Yes & Yes  \\ 
  & Cochlear filter bank type\footnotemark[3] & GTF & GTF  & GTF    & GTF & DRNL\\ \hline
3 /& Half-wave rectification & `ihc\_breebaart' & `ihc\_dau'\footnotemark[1]  & `ihc\_dau'\footnotemark[1]  & `ihc\_breebaart'\footnotemark[1] & `ihc\_dau'\footnotemark[1] \\
  & Expansion         & No &  No           & No     & No  & Yes  \\
  & LPF, $f_c$~Hz (filter order) & 770 (5) & 1000 (1) & 1000 (1) & 770 (5) & 1000 (1) \\ \hline
4 /& Adaptation loops  & `adt\_osses2020' & `adt\_dau'\footnotemark[1] & `adt\_dau'\footnotemark[1] & `adt\_breebaart'\footnotemark[1] & `adt\_dau'\footnotemark[1] \\
  & Set of constants $\tau$\footnotemark[4]  & A &  A  &  A         & B & A    \\
  & Limitation (factor) & Yes (5) & Yes (10) & Yes (10)  & No ($+\infty$) & Yes (10) \\ \hline
5 /& Modulation filter bank & `mfb\_jepsen2008' & `mfb\_dau1997'\footnotemark[1] & `mfb\_verhey1999' & default\footnotemark[2] & `mfb\_jepsen2008'\footnotemark[1] \\
  & Number of modulation filters & 12 & 12     & 12     & 1 & 12   \\ 
  & 150-Hz LPF        & Yes & No           & No      & No & Yes \\  
  & Limited to filters with mf$_c<f_c/4$ & Yes & No & Yes & No & Yes \\
  & Scaling factor ($1/\sqrt{2}$) & Yes & No & No & No & Yes \\ \hline\hline
\end{tabular}
}
\footnotetext[1]{Default AMT flag for the corresponding preprocessing model (*\_preproc.m).}
\footnotetext[2]{No flag is used within AMT as this is the only possible stage configuration for the corresponding preprocessing model.}
\footnotetext[3]{Cochlear filter bank types: GTF = Gammatone (linear) filter bank; DRNL = Dual-resonance non-linear filter bank.}
\footnotetext[4]{Set A of time constants: $\tau$=5, 50, 129, 253, 500~ms; Set B, linearly spaced between 5 and 500~ms: $\tau$=5, 128.75, 252.5, 376.25, 500~ms.}
\label{tab:models-param}
\end{table*}



The implementation of peripheral stages of the PEMO model (Stages 1-5, Fig.~\ref{fig:auditory-model}) as used in this study is available within the AMT toolbox (v0.10) for MATLAB \cite{Soendergaard2013}. This implementation required the attachment of the outer- and middle-ear processing (Stage 1) and the extension of configuration parameters within the \textsf{dau1997\_preproc} routine. The model parameters and its variants are summarized in Table~\ref{tab:models-param}.

Figures~\ref{fig:energy-per-band} and \ref{fig:adapt-loops-tones} of this paper can be reproduced in AMT by using \textsf{exp\_dau1997(`fig4\_osses2020')} and \textsf{exp\_dau1997(`fig14\_osses2020')}, respectively.

\section{Adaptation loops}
\label{app:adaptation}

The adaptation loops stage is a digital feedback structure included in the PEMO model and in other variants of this model (see Sec.~\ref{sec:model-description}). The adaptation loops stage simulates the adaptive properties of the hearing system \cite{Westerman1984, Kohlrausch1992}. The most relevant properties of the adaptation loops are highlighted in this appendix, which were partly collected from the original implementation descriptions \cite{Pueschel1988, Muenkner1993, Dau1996a} and were further explored in our study to justify the use of a more severe limitation of the onset overshoot (limiter factor lim=5) with respect to the literature (no limiter factor or lim=10).

\subsection{Adaptation and use of the RC analogy}

\begin{figure}[b]
	\centering
	\includegraphics[width=0.48\textwidth]{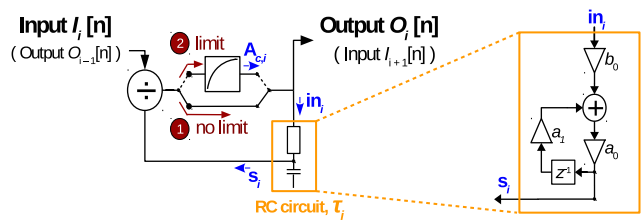}
	\caption{Structure for the adaptation loop~$i$. Signal path ``1'' corresponds to the implementation without output limitation \cite{Pueschel1988}. Signal path ``2'' includes a logistic growth compressor that limits the overshoot responses of the system \cite{Muenkner1993}. These structures are described in the text. The inset shows the lattice structure of the first-order IIR filter between in$_i$ and $s_i$.}
	\label{fig:adapt-loops}
\end{figure}

The adaptation loops receive signals after the cochlear band-pass filtering and subsequent inner hair cell processing (after Stage~3 in the PEMO model, Fig.~\ref{fig:auditory-model}). The $i$-th adaptation loop (with $i$ from 1 to 5) is implemented as a first-order IIR filter (time constants $\tau$=5, 50, 129, 253, 500~ms) that corresponds to a resistor-capacitor (RC) circuit (Fig.~\ref{fig:adapt-loops}). This digital structure acts as a low-pass filter between node in$_i$ and output $s_i$. Output $s_i$ represents the charging state of the filter and ranges between an initial charging state $s_0$ and 1, with shorter or longer time constants producing a faster or slower charge/discharge of the circuit, respectively. Furthermore, an uncharged RC circuit amplifies the incoming signal and a fully charged circuit does not alter the amplitude of the incoming signal. This results in a nearly logarithmic transformation of stationary input signals \cite[see Fig.~\ref{fig:adaptIO}\textbf{B}, and also,][their Fig.~3]{Dau1996a}. The initial charging state $s_0$ is defined by the minimum input level lvl\textsubscript{min} to the first adaptation loop, which is set to an amplitude of $1\cdot 10^{-5}$ (0~dB SPL for a full scale convention of 100~dB): 
\begin{equation}
	s_{0,i} = \left(1/a_0\right)\cdot\sqrt[2^i]{\mbox{lvl\textsubscript{min}}} \hspace{20pt} \mbox{with }i=1,2, \cdots ,5 
\end{equation}



\noindent where $a_0$=1, is one of the coefficients of the difference equation that characterizes the IIR filter between in$_i$ and $s_i$. The full  difference equation is given by: 
\begin{equation}
	a_0 \cdot s_i[n]-a_{1,i} \cdot s_i[n-1] = b_{0,i} \cdot \mbox{in}_i[n]
	\label{eq:a0s0}
\end{equation}

\noindent where $a_0$=$1$, $a_{1,i}$=$\exp{\left( -1/\left(\tau_i \cdot f_s\right) \right)}$, and $b_{0,i}$=$1-a_{1,i}$. The output $O[n]$, after the five adaptation loops, is finally scaled in a way that outputs with an amplitude of $1$ (100~dB steady inputs) are mapped to 100~MU and outputs with an amplitude of $s_{0,5}$=$0.6978$ (Eq.~\ref{eq:a0s0}) are mapped to 0~MU:
\begin{equation}
	\Psi[n] = 100 \cdot \left(O[n]-s_{0,5}\right)/\left(1-s_{0,5}\right)
	\label{eq:Psi}
\end{equation}


\subsection{Overshoot limitation}

The strong onset response obtained with the adaptation loops structure of Fig.~\ref{fig:adapt-loops}, ``no-limit path,'' motivated the introduction of a limiter at the output of each individual loop \cite{Muenkner1993}. The so-called overshoot limitation is implemented as a compressor with a ratio that follows a logistic growth described by:
\begin{equation}
	A_{c,i} = 
	\begin{cases} 
		\mbox{in}_i & \mbox{if }\mbox{in}_i\leq 1\\ 
		\frac{2 \cdot C_i}{1+\exp{\left[ \frac{2}{C_i} \cdot \left( 1-\mbox{in}_i\right) \right]}}+(1-C_i) & \mbox{if }\mbox{in}_i > 1
	\end{cases}
	\label{eq:limitation}
\end{equation}

\noindent This equation implements a compression to the input in$_i$, obtaining an output $A_{c,i}$ (Fig.~\ref{fig:adapt-loops}, signal path ``2''). The compressor has a threshold of $1$ (non-normalized amplitude) and a limiter threshold \textsf{thres}\textsubscript{lim,$i$} that depends on an arbitrary limiter factor lim and on the initial charge of each adaptation loop. For convenience, a constant $C_i$ (to be used in Eq.~\ref{eq:limitation}) is also defined: 
\begin{eqnarray}
	\mbox{thres\textsubscript{lim,$i$}} = (1-s^2_{0,i}) \cdot \mbox{lim} \nonumber \\
	C_i = \mbox{thres\textsubscript{lim,$i$}}-1
	\label{eq:thres-lim}
\end{eqnarray}

For a limiter factor lim=10, as suggested by \citeauthor{Muenkner1993}, the limiter thres\textsubscript{lim,1} is equal to 10 ($C_1$=9, $s_{0,1}$=0.0032), hence producing outputs of loop 1 of maximally 10 times the input amplitude in$_1$, but that due to the subsequent compression in the remaining 4 loops, results in a maximum possible output amplitude thres\textsubscript{lim,5} of 5.1 ($C_5$=4.1, $s_{0,5}$=0.6978). For a limiter factor lim=5, as suggested in the current study, thres\textsubscript{lim,1} is equal to 5 ($C_1$=4, $s_{0,1}$=0.0032) resulting in a maximum possible output thres\textsubscript{lim,5} of 2.6 (with $C_5$=1.6, $s_{0,5}$=0.6978). The effect of the adaptation loops on normalized outputs $\Psi[n]$ (Eq.~\ref{eq:Psi}) are illustrated in Figs.~\ref{fig:adapt-loops-tones} and \ref{fig:adaptIO}. 

\subsection{Input-output characteristic}

The effect of the adaptation loops on normalized outputs $\Psi[n]$ is shown for pure tones with a carrier frequency of 4000~Hz (duration of 300~ms, 2.5-ms up/down ramps) using no limiter factor (equivalent to lim$\to+\infty$), lim=10, and lim=5. This stimulus choice is similar to the sounds employed by \citet[][see their Fig.~6]{Westerman1984} in recordings of auditory nerve responses of the Mongolian gerbil, which served as a reference for the development of the overshoot limitation scheme \cite{Muenkner1993}. The stimuli shown here have, however, levels between 10 and 100~dB SPL (\citeauthor{Westerman1984} used stimuli with levels up to 40~dB). 

\begin{figure}
	\centering
	\includegraphics[width=0.44\textwidth]{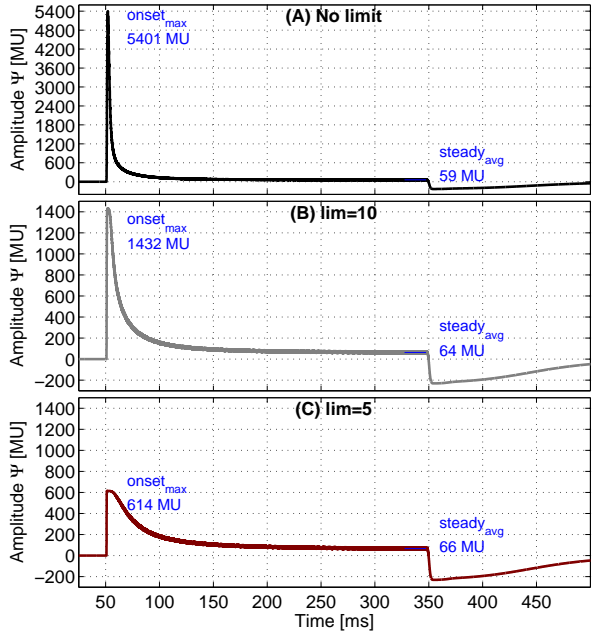}
	\caption{Adaptation loops output $\Psi$, expressed in MU, for a 4000-Hz sine tone with a level of 70~dB~SPL using (\textbf{A}) non-limited amplitudes, and for amplitudes limited with factors (\textbf{B}) lim=10, and (\textbf{C}) lim=5, as adopted in our study. For clarity, the amplitude scale in (\textbf{A}) is not to scale with (\textbf{B},\textbf{C}) Further details are given in the text.}
	\label{fig:adapt-loops-tones}
\end{figure}

We first show in Fig.~\ref{fig:adapt-loops-tones}\textbf{A}--\textbf{C} the effect of the different limiter factors for a pure tone of 70 dB for lim$\to+\infty$ (no factor), lim=10, and  lim=5, respectively. Maximum onset (onset\textsubscript{max}) and average steady-state (steady\textsubscript{avg}) amplitudes were calculated, with the later being obtained from the average of the last 20~ms of the $\Psi$ amplitudes before the signal offset. The onset amplitude onset\textsubscript{max} in (\textbf{A}) was 5401~MU, which is 91.5 times the steady-state response  steady\textsubscript{avg} of 59~MU. Such overshoot supports the need for the overshoot limitation that produced an onset\textsubscript{max} of 1432~MU (22.4 times the steady\textsubscript{avg} value of 64~MU), and 614~MU (9.3 times the steady\textsubscript{avg} value of 66 MU), for (\textbf{B}) lim=10 and (\textbf{C}) lim=5, respectively. This figure shows that lim=10 does not produce an overshoot limitation of 10 times the steady-state response as claimed in the literature. That is actually the case if only one adaptation loop were to be used (Eq.~\ref{eq:thres-lim}). 

\begin{figure}
	\centering
	\includegraphics[width=0.44\textwidth,]{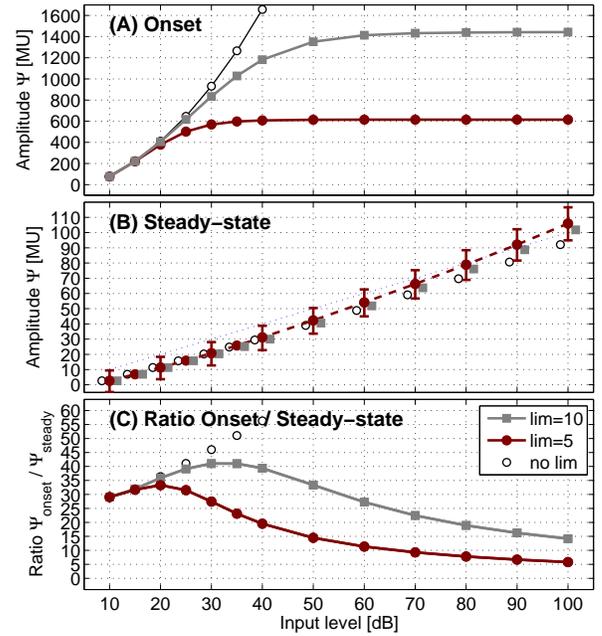}
	\caption{Input-output characteristic function of the adaptation loops for (\textbf{A}) onset and (\textbf{B}) steady-state responses to 4000-Hz pure tones using limiter factors lim=5 (maroon circles) and lim=10 (gray squares). The error bars in (\textbf{B}) indicate $\Psi\textsubscript{min}$ and $\Psi\textsubscript{max}$ amplitudes, in the 20-ms integration period and show that the temporal fine structure is not completely removed by the 770-Hz LPF (IHC processing). Non-limited responses are shown as a reference (black lines). In panel (\textbf{C}) the onset to steady-state ratio is shown.}
	\label{fig:adaptIO}
\end{figure}

%
%

The input-output characteristic functions for the onset and steady-state responses to the 4000-Hz test tones are shown in Fig.~\ref{fig:adaptIO}\textbf{A} and \textbf{B}, respectively. The tones were presented at levels between 10 and 100~dB SPL in steps of 10 dB for the three limitation configurations. The onset to steady-state ratio is shown in Fig.~\ref{fig:adaptIO}\textbf{C}.

In Fig.~\ref{fig:adaptIO}\textbf{A}, the onset of the non-limited responses continue to grow outside the range indicated in the figure, while for the responses with lim=10 (gray squares) the onsets are (1) almost unaffected for input levels up to 20~dB, (2) compressed for levels between 20 and 50 dB, and (3) limited to approximately 1442 MU, with a minimum $\Psi\textsubscript{onset}/\Psi\textsubscript{onset}$ ratio equal to 14.2 (Fig.~\ref{fig:adaptIO}\textbf{C}). For responses with lim=5, the compressing range extends from 20 to 35~dB, with higher input levels being limited to 614~MU (minimum $\Psi\textsubscript{onset}/\Psi\textsubscript{onset}$ ratio of 5.8).

The steady-state responses in Fig.~\ref{fig:adaptIO}\textbf{B} show that they do not change  considerably for the different adaptation loops configurations. In addition, the error bars shown in the figure (for clarity only for the responses with lim=5) indicate maximum and minimum $\Psi$ amplitudes over the 20-ms integration period. This indicates that not all phase information was removed by the fifth-order 770-Hz LPF (IHC processing), i.e., there is still some temporal fine structure present in these $\Psi$ responses. 

\subsection{Implications of the onset limitation in the current study}

The ratios between onset and steady-state responses $\Psi\textsubscript{onset}/\Psi\textsubscript{onset}$ shown in Fig.~\ref{fig:adaptIO}\textbf{B} show values that are overall higher than the target ratio of 10 observed by \citet{Westerman1984}. Adopting a factor lim=5, produced $\Psi\textsubscript{onset}/\Psi\textsubscript{onset}$ ratios that were closer to that target. The piano sounds investigated in this paper have prominent onset characteristics, which required a stronger overshoot limitation (lim=5), which would have otherwise lead to unsuccessful simulation results, even though this choice did not affect considerably other psychoacoustic tasks \cite[some of them taken from][]{Jepsen2008} that we simulated to validate the custom PEMO model configuration we used \cite[App.~D of][not shown in this paper]{Osses2018}.

\end{document}